\documentclass[12pt,a4paper,dvips]{article}
\usepackage{amssymb,amsmath,graphicx}
\usepackage{cite,mcite}
\usepackage{a4p}
\usepackage{ifthen}
\usepackage{l3_title}
\usepackage{Lep}
\usepackage{higgs_input}
\Lep{2}
\journalname{Phys. Lett. B}
\date{August 2, 2002}
\preprint{2002-066}       

\newlength{\capindent}
\newlength{\capwidth}
\newlength{\figwidth}
\newcommand{\icaption}[2][!*!,!]{\hspace*{\capindent}%
  \begin{minipage}{\capwidth}
    \ifthenelse{\equal{#1}{!*!,!}}%
      {\caption{#2}}%
      {\caption[#1]{#2}}
  \end{minipage}}

\newcommand{\mh}{m_\mathrm{h}}

\newcommand{\mA}{m_\mathrm{A}}
\newcommand{\mZ}{m_\mathrm{Z}}
\def\ra{\rightarrow}

\begin{document}
\begin{titlepage}
\title{ Search for Neutral Higgs Bosons of the Minimal Supersymmetric
    Standard Model in $\boldsymbol{\epem}$ Interactions at
    $\boldsymbol{\sqrt{s}}$ up to 209 $\boldsymbol{\GeV}$
}

\author{L3 Collaboration}


\begin{abstract}
  A search for the lightest neutral CP-even and neutral CP-odd
  Higgs bosons of the Minimal Supersymmetric Standard Model is
  performed using 216.6\pb of data collected with the
  L3 detector at LEP at 
  centre-of-mass energies between 203 and 209\GeV. 
  No indication of a signal is found. Including our 
  results from lower centre-of-mass energies,  
  lower limits on the Higgs boson masses are
  set as a function of \tanb for several 
  scenarios. For \tanb greater than 0.7 they are
  $\mh >$ 84.5\GeV and 
  $\mA >$ 86.3\GeV at 95\% confidence level.
\end{abstract}

\submitted

\end{titlepage}

%
\section{Introduction}
\label{sec:intro}

In the Minimal Supersymmetric Standard
Model (MSSM)~\cite{mssm_1}
two doublets of complex scalar 
fields are required 
to generate the masses of gauge bosons
and fermions.
The neutral Higgs sector of the MSSM comprises three physical states:
two CP-even Higgs bosons, 
the lighter of which is denoted as \h and the heavier as \bigH,
and a neutral CP-odd boson, \A.

The two most important production mechanisms of the 
light neutral Higgs boson in \epem collisions are: 
\begin{equation}
\mathrm{
e^+e^-\rightarrow hZ, 
}
\label{eq:higgstrahlung}
\end{equation}
\begin{equation} 
\mathrm{
e^+e^-\rightarrow hA, 
}
\label{eq:higgspairprod}
\end{equation}
with tree level cross sections that are related to the  
Standard Model Higgs-strahlung cross section, $\mathrm{\sigma_{HZ}^{SM}}$, 
as~\cite{higgs_hunters}:
\begin{equation}
\mathrm{
\sigma_{hZ}=\sin^2(\beta-\alpha)\sigma_{HZ}^{SM},
}
\end{equation}
\begin{equation}
\mathrm{
\sigma_{hA}=\cos^2(\beta-\alpha)\tilde{\lambda}\sigma_{HZ}^{SM},
}
\end{equation}
where \tanb is the ratio 
of the vacuum expectation values of the two Higgs 
doublets, $\mathrm{\alpha}$ is
the mixing angle in the CP-even Higgs boson sector
and $\mathrm{\tilde{\lambda}}$ is the p-wave 
suppression factor depending on the Higgs boson 
masses, $\mh$ and $\mA$, and the centre-of-mass energy $\sqrt{s}$. 
For some choices of the MSSM parameters,
the \bigH boson can also be produced
via the Higgs-strahlung process
\begin{equation}
\mathrm{
e^+e^-\rightarrow HZ,
}
\label{eq:bighiggstrahlung}
\end{equation}
with a cross section suppressed by the factor $\mathrm{\cos^2(\beta-\alpha)}$
relative to $\mathrm{\sigma_{HZ}^{SM}}$. 

For most of the MSSM parameter space considered, 
the neutral Higgs bosons are predicted to decay dominantly into 
$\mathrm{b\bar{b}}$ and $\mathrm{\tau^+\tau^-}$. 
However, in certain parameter regions, other decays like 
$\mathrm{h\rightarrow AA}$ and 
$\mathrm{A\rightarrow c\bar{c}}$ become important.

The search for the neutral Higgs bosons is 
performed in the framework of the constrained MSSM 
with seven free parameters. These are
the universal sfermion mass parameter, $\mathrm{M_{SUSY}}$,
the common Higgs-squark trilinear coupling, $\cal{A}$,
the supersymmetric Higgs mass parameter, $\mu$,
the SU(2) gaugino mass parameter, $\mathrm{M_2}$,
the gluino mass parameter, $\mathrm{M_3}$, $\mA$ and \tanb. 
The mass of the top quark is taken to be 
174.3\GeV~\cite{top_mass}. 

Previous searches for the neutral Higgs bosons were reported 
by L3~\cite{l3_2000,l3_1999} and other
experiments~\cite{opal_mssm}.
In this Letter, we present the results of the search for the h and A bosons
using data collected with the L3
detector~\cite{l3_det} in the year 2000. 
In comparison to previous analyses~\cite{l3_2000} the 
performance in the four-jet channel
is improved by using a new 
likelihood based analysis. In addition, the six-jet final state 
resulting from the $\mathrm{hZ\ra AA q\bar{q}}$ topology
is investigated.

%
\section{Benchmark Scenarios}
\label{sec:CMSSM}

Due to the large number of parameters remaining 
in the constrained MSSM, we focus on three specific
parameter settings as suggested in Reference~\citen{mssm_benchmark}.
These benchmark settings are denoted as the ``$\mh-$max'', ``no mixing''
and ``large$-\mu$'' scenarios, with the corresponding MSSM parameter values
detailed in Table~\ref{tab:mssm_param}. 
The first two scenarios take into account radiative corrections to $\mh$
computed within a two-loop diagrammatic 
approach~\cite{heinemeyer_1}
and differ only by the value of $\mathrm{X_t\equiv{\cal{A}}-\mu\cot\beta}$,
which governs the mixing in the scalar 
top sector. The ``$\mh-$max'' scenario is designed to 
extend the search to the maximal theoretical bound on $\mh$
for any value of $\tanb$ and leads to rather conservative
constraints on $\tanb$. The ``no mixing'' scenario 
corresponds to vanishing mixing in the scalar top sector 
and is more favorable to LEP searches.
After fixing the parameters $\mathrm{M_{SUSY}}$, $\mathrm{M_2}$,
$\mathrm{\mu}$, $\mathrm{X_t}$ and $\mathrm{M_3}$, 
a scan over the remaining parameters \tanb and $\mA$ 
is performed in the range
0.4 $\le$ \tanb $\le$ 30 and 10\GeV $\le \mA \le$ 1\TeV.
The widths of the \h and \A bosons are assumed to be
smaller than the experimental mass resolutions.
As this holds in these two scenarios only for \tanb $\le$ 30, higher
values of \tanb are not considered.

In the ``large$-\mu$'' scenario 
the upper theoretical bound on 
$\mh$ is slightly less than 108\GeV, thus in 
the LEP reach.
However, for some choices of \tanb and $\mA$, 
the Higgs boson pair production~(\ref{eq:higgspairprod})
is kinematically inaccessible and the Higgs-strahlung 
process~(\ref{eq:higgstrahlung}) is strongly 
suppressed due to small values of $\mathrm{\sin^2(\beta-\alpha)}$.
Sensitivity in these regions can be recovered by 
exploiting the \bigH boson production via the
high cross section Higgs-strahlung process~(\ref{eq:bighiggstrahlung}).
The scan is performed over $\mA$ from 10\GeV
to 400\GeV and over \tanb between 0.7 and 50. 
Within this scenario, the assumption that the widths of the 
Higgs bosons are small compared to experimental mass resolutions 
is valid for \tanb up to 50.  
For the interpretation of the data within this ``large$-\mu$'' scenario,
the improved one-loop renormalisation group 
calculations~\cite{carena_1} are used.

%
\section{Data and Monte Carlo Samples}

The data recorded at centre-of-mass
energies between 203 and 209$\GeV$ are grouped 
into three data sets with
effective centre-of-mass energies of 204.3, 206.1 and 208.0\GeV,
corresponding to integrated
luminosities of 26.0, 181.9 and 8.7\pb. 
The results obtained from this data are combined with the results from
integrated luminosities of 233.2\pb at 
192\GeV$<\mathrm{\sqrt{s}}<$ 202$\GeV$~\cite{l3_2000} 
and 176.4\pb at $\mathrm{\sqrt{s}} =$ 189$\GeV$~\cite{l3_1999}.

The cross sections of processes~(\ref{eq:higgstrahlung}), 
(\ref{eq:higgspairprod}) and~(\ref{eq:bighiggstrahlung}) 
and the decay branching fractions of h, H and A are calculated using
the HZHA generator~\cite{hzha}. For efficiency studies, Monte
Carlo samples of 2000 Higgs events are generated for each mass hypothesis 
in each search channel using PYTHIA~\cite{pythia} and HZHA.  
For the \hA~samples, $\mh$ and $\mA$ range
from 50 to 105\GeV in steps of 5\GeV. For the \hZ~samples, $\mh$ is chosen
in steps of 5\GeV from 60 to 100\GeV and in steps of 1\GeV from 100 
to 120\GeV. 
For background studies, the following Monte Carlo programs are used:
KK2f~\cite{kk2f} (\epemtoqqg), PYTHIA (\epemtoZZ and
$\epem\!\rightarrow\!\Z\epem$), KORALW~\cite{koralw} 
(\epemtoWW) and KORALZ~\cite{koralz} (\epemtotautau).
Hadron production in two-photon interactions is simulated with PYTHIA 
and PHOJET~\cite{phojet}.
EXCALIBUR~\cite{excalibur} is used for other four fermion final states. 
The number of simulated 
events for the most important background channels is more than 100
times the number of expected events.  

The L3 detector response is simulated using the GEANT 
program~\cite{geant}, which models the effects of energy
loss, multiple scattering and showering in the detector. The GHEISHA
program~\cite{gheisha} is used to simulate hadronic interactions in
the detector. Time dependent
detector inefficiencies, monitored during data taking,
are also taken into account.

%
\section{Analysis Procedures}

For the hA production, the decay modes considered are:
$\mathrm{hA\rightarrow \mathrm{b\bar{b}b\bar{b}}}$, 
$\mathrm{hA\rightarrow \mathrm{b\bar{b}\tau^+\tau^-}}$ and 
$\mathrm{hA\rightarrow \mathrm{\tau^{+}\tau^{-}b\bar{b}}}$.
For the hZ or HZ production, the 
event topologies considered are: $\mathrm{q\bar{q}q^\prime\bar{q}^\prime}$,
$\mathrm{q\bar{q}\nu\bar{\nu}}$, 
$\mathrm{q\bar{q}\ell^+\ell^-(\ell=e,\mu,\tau)}$ and
$\mathrm{\tau^+\tau^- q\bar{q}}$. To cover the MSSM parameter 
regions where the decay $\mathrm{h\rightarrow AA}$ becomes important,  
the $\mathrm{hZ\ra AA q\bar{q}\ra q\bar{q}q^\prime\bar{q}^\prime q^{\prime\prime}\bar{q}^{\prime \prime}}$ channel is studied as well.
Searches in channels with 
decays of the \h boson into quarks are optimised for 
the dominant $\mathrm{h\rightarrow b\bar{b}}$ decay channel.
The analyses of the $\mathrm{q\bar{q}\nu\bar{\nu}}$ and 
$\mathrm{q\bar{q}\ell^+\ell^-(\ell=e,\mu)}$ 
final states are the same 
as those used in the Standard Model Higgs search~\cite{l3_sm_paper}.

%
\subsection{The \boldmath{$\rm{hZ\ra b\bar{b}q\bar{q}}$} and 
\boldmath{$\rm{hA\ra b\bar{b}b\bar{b}}$} Analyses}
\label{sec:bbbb}

The signature of the $\mathrm{hZ\ra b\bar{b}q\bar{q}}$
final state is four high-multiplicity
hadronic jets and the presence of b hadrons,
in particular in the jets expected to stem from the Higgs boson.
The invariant mass of the jets supposed to originate from 
the Z boson must be compatible, within mass resolution, with its mass, $\mZ$.
The $\mathrm{hA\ra b\bar{b}b\bar{b}}$ search topology is characterised by 
four high multiplicity hadronic jets originating 
from b-quarks. The main backgrounds arise from 
$\mathrm{q\bar{q}(\gamma)}$ final states
and hadronic decays of W-pairs and Z-pairs. 

Initially, a common preselection 
for both the $\mathrm{hZ\ra b\bar{b}q\bar{q}}$ and 
$\mathrm{hA\ra b\bar{b}b\bar{b}}$ channels is applied, followed 
by a kinematic classification of events into hZ or hA analysis branches.
Selection criteria optimised for each branch are used and a final discriminant 
specific to each branch is constructed. 

The preselection criteria used are described in Reference~\citen{l3_sm_paper}.
Events passing the preselection are forced into a four-jet topology
using the Durham algorithm~\cite{DURHAM} and a kinematic fit imposing
energy and momentum conservation (4C fit) is performed. 
Each event is tested for its compatibility with the hZ and hA
production hypotheses, exploiting dijet invariant masses.
There are three possibilities for pairing jets in a four-jet event.
For each pairing,  
$\mathrm{\chi^2}$ values are calculated
for the hypotheses of hZ and hA production:
\begin{equation}
\mathrm{
\chi^2_{hZ}=
\frac{\left(\Sigma_i-{\it{m}}_h-{\it{m}}_Z\right)^2}{\sigma_\Sigma^2}+
\frac{\left(\Delta_i-|{\it{m}}_h-{\it{m}}_Z|\right)^2}{\sigma_\Delta^2},
}
\label{eq:chi2hz}
\end{equation}
\begin{equation}
\mathrm{
\chi^2_{hA}=
\frac{\left(\Sigma_i-{\it{m}}_h-{\it{m}}_A\right)^2}{\sigma_\Sigma^2}+
\frac{\left(\Delta_i-|{\it{m}}_h-{\it{m}}_A|\right)^2}{\sigma_\Delta^2}.
}
\end{equation}
In these expressions $\mathrm{\Sigma_i}$ and 
$\mathrm{\Delta_i}$ are the dijet mass sum and dijet mass difference
of the i-th pairing, and 
$\mathrm{\sigma_\Sigma}$ and
$\mathrm{\sigma_\Delta}$ are the dijet mass sum and 
dijet mass difference resolution functions. They are found 
to be almost independent of the dijet masses
for production well above threshold and are estimated from Monte Carlo 
to be $\mathrm{\sigma_\Sigma=4\hspace{1mm}GeV}$ and 
$\mathrm{\sigma_\Delta=18\hspace{1mm}GeV}$.
Close to the kinematic threshold
they strongly depend on the dijet mass sum.
For each hypothesis, the jet pairing with the smallest $\mathrm{\chi^2}$ is 
chosen and the probability $\mathrm{P(\chi^2)}$ is calculated.

Events are assigned either to the hZ or hA analysis branch by means 
of a binned likelihood~\cite{l3_sm_paper}, 
$\mathrm{L_{hZ}^c\equiv 1-L_{hA}^c}$,
constructed from the following variables: the
$\chi^2$ probabilities, $\mathrm{P(\chi^2_{hZ})}$ and
$\mathrm{P(\chi^2_{hA})}$, the Higgs boson 
production angle, $\mathrm{|\cos\Theta|}$,
the number of charged tracks, $\mathrm{N_{trk}}$,
the global event b-tag, $\mathrm{B_{tag}}$, and 
the maximum triple jet boost, $\mathrm{\gamma_{triple}}$~\cite{l3_sm_paper}. 
Events are assigned to the hZ branch if 
$\mathrm{L_{hZ}^c>0.5}$, or to the hA branch otherwise. 

The selection in both branches 
proceeds in the same way. Events are rejected 
if $\mathrm{P(\chi^2)}<1\%$. High b-tag events are accepted
and a selection likelihood is then constructed
to separate the signal from the $\mathrm{q\bar{q}(\gamma)}$ 
final states and hadronic decays of W-pairs and Z-pairs, using 
$\mathrm{N_{trk}}$, $\mathrm{B_{tag}}$, 
$\mathrm{\gamma_{triple}}$, $\mathrm{|\cos\Theta|}$ and 
$\mathrm{logY_{34}}$, where $\mathrm{Y_{34}}$ is 
the jet resolution parameter for which the event topology changes from 
three to four jets.
A final discriminant 
is constructed for events passing an optimised selection 
likelihood cut. The 
optimisation is based on the analysis performance at  
($\mA$,$\mh$) values close to the expected sensitivity of 
the L3 combined search.
The final discriminants are built from 
individual b-tag variables of the four jets and 
$\mathrm{P(\chi^2)}$.
In the hZ branch, an event category variable is also used. This variable is constructed from rankings of b-tag variables of the two jets assigned 
to the Higgs boson, as described in Reference~\citen{l3_sm_paper}.  
The final discriminant in the hA branch also includes  
the corresponding selection likelihood. 

\begin{figure}
\begin{tabular}{cc}
\hspace{-7mm}
\includegraphics*[width=0.5\textwidth]{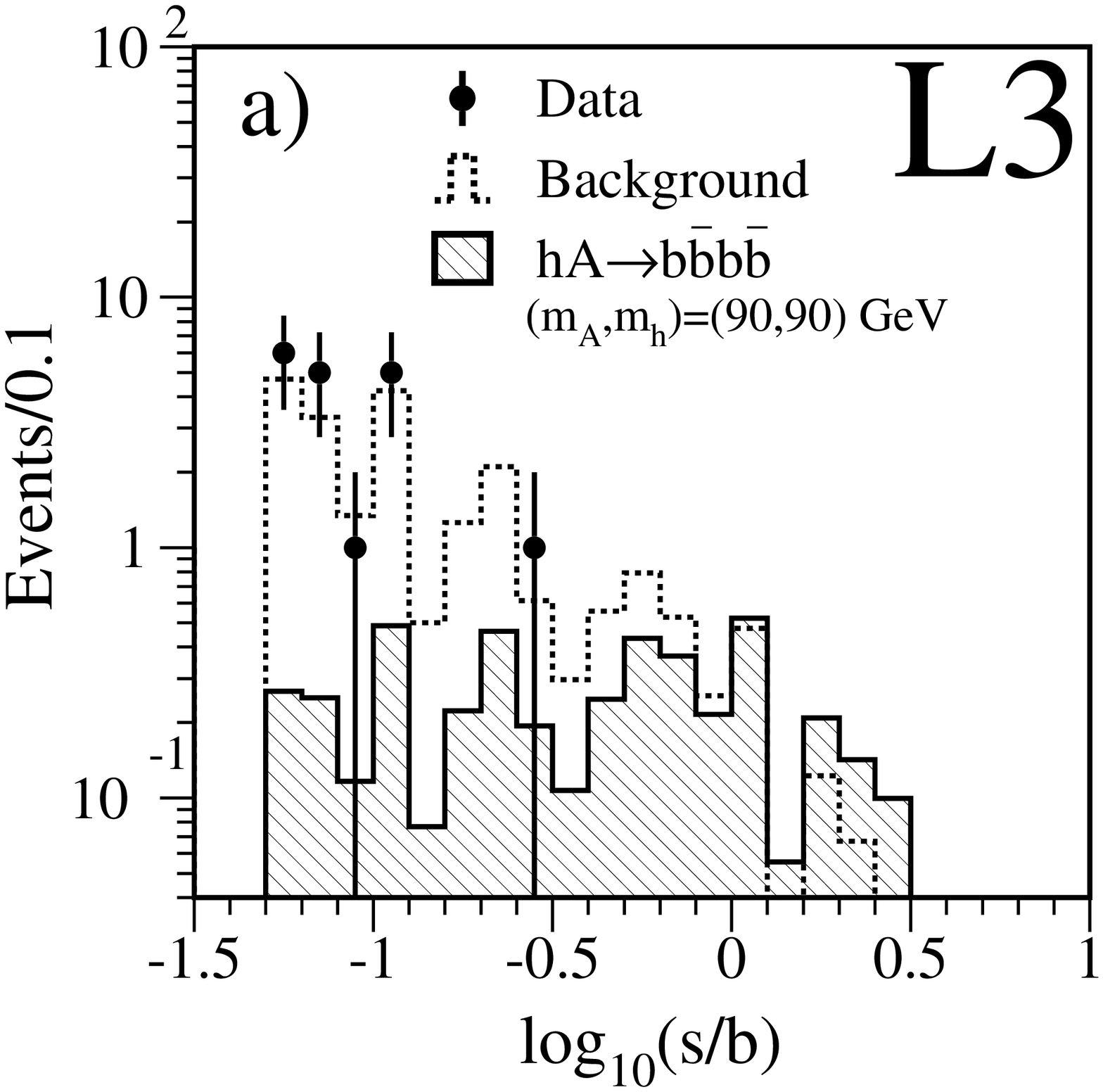} &
\hspace{-0mm}
\includegraphics*[width=0.5\textwidth]{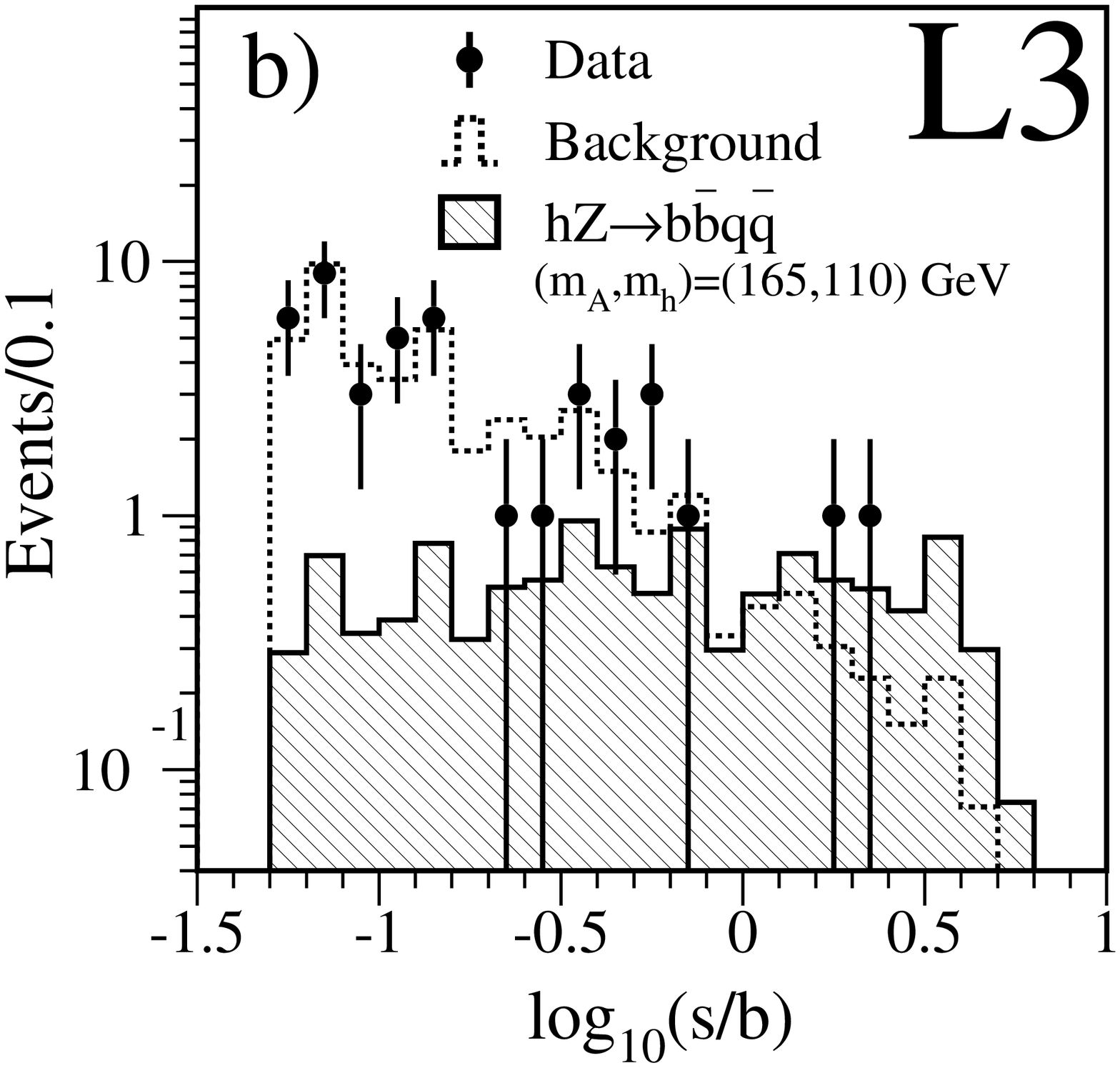} \\
\hspace{-7mm}
\includegraphics*[width=0.5\textwidth]{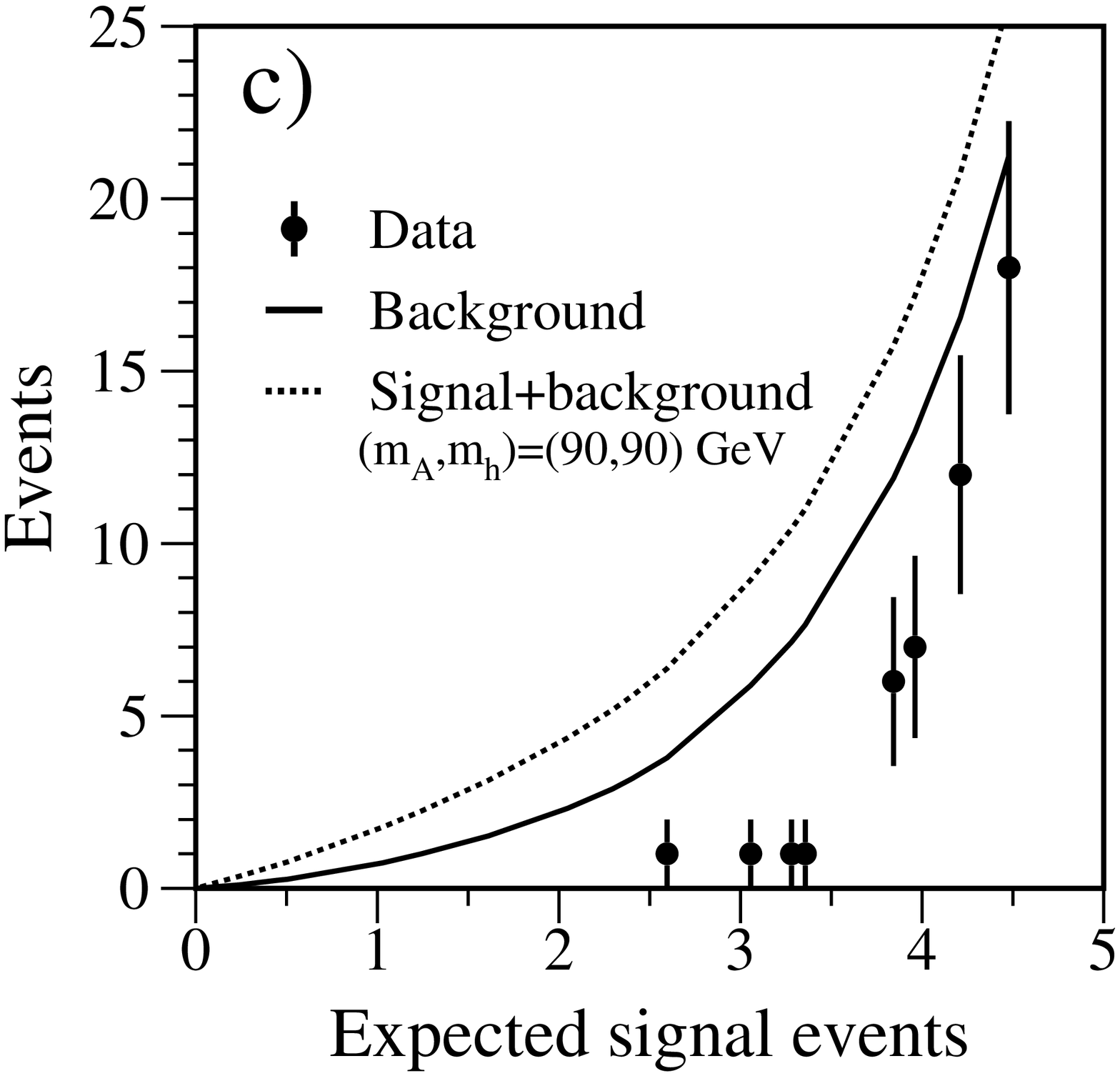} &
\hspace{-0mm}
\includegraphics*[width=0.5\textwidth]{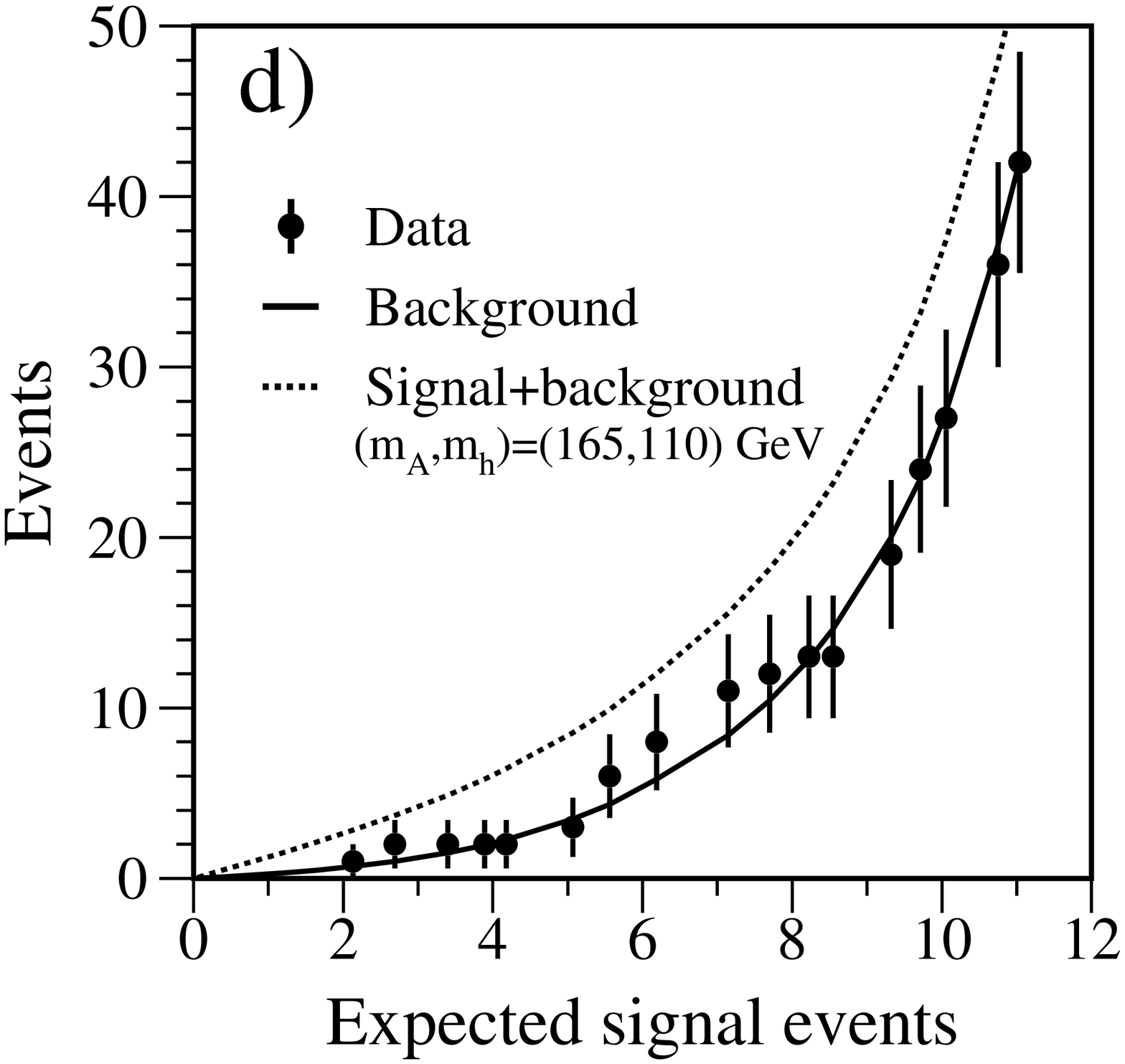} \\
\end{tabular}
\caption{ The distribution of data, expected 
background and expected signal events as a function of 
the logarithm of the signal-to-background ratio,
$\mathrm{log_{10}(s/b)}$, in the four-jet channel for the Higgs boson mass 
hypotheses a) ($\mA$,$\mh$) = (90,90) GeV and 
b) ($\mA$,$\mh$) = (165,110) GeV.
Integrated distributions of data and expected background events as 
a function of the expected signal are shown in  
c) and d).
}
\label{fig:bbbb}
\end{figure}

Table~\ref{tab:bbbb} reports
the number of data, expected background and expected signal
events selected at different stages of the analysis 
for two representative
Higgs boson mass hypotheses in the ``$\mh-$max'' scenario, 
($\mA$,$\mh$) = (90,90)\GeV at \tanb = 25 
and ($\mA$,$\mh$) = (165,110)\GeV at \tanb = 3.
For these hypotheses the distribution of the final discriminant
in terms of a signal-to-background ratio 
is presented in Figure~\ref{fig:bbbb}.
The data are compatible with the background expectation.

%
\subsection{The \boldmath{$\rm{hZ\ra b\bar{b}\tau^+\tau^-,\tau^+\tau^- q\bar{q}}$} and \boldmath{$\rm{hA\ra b\bar{b}\tau^+\tau^-}$} Analyses}
\label{sec:bbtt}

The signatures of $\mathrm{hZ\ra b\bar{b}\tau^+\tau^-,\tau^+\tau^- q\bar{q}}$
and $\mathrm{hA\ra b\bar{b}\tau^+\tau^-}$
\footnote{Both of the decay modes 
(h$\rightarrow \mathrm{b}\mathrm{\bar b}$, A$\rightarrow\tau^+\tau^-$)
and
(h$\rightarrow\tau^+\tau^-$, A$\rightarrow \mathrm{b}\mathrm{\bar b}$) 
are considered.} 
final states are a pair of taus accompanied by two hadronic jets.  
For each of the channels hA and hZ an analysis is optimised
based either on the tau identification or on
the event topology by requiring four jets 
with two of them being narrow and of low multiplicity.
The main background comes from W-pair decays containing taus.
The analysis is similar to the one
used in previous searches~\cite{l3_2000}.
The selection is optimised for lower Higgs boson masses
by applying looser cuts on the opening angles of
the jets and tau pairs compared to the Standard Model Higgs 
search~\cite{l3_sm_paper}. 
The invariant mass of the tau pair, $m_{\rm{\tau^+\tau^-}}$, and
of the hadronic jets, $m_{\rm{q\bar{q}}}$,
must be between 25\GeV and
125\GeV.  
The ratio between
the total energy deposited in the detector, $\mathrm{E_{vis}}$,
and $\sqrt s$ must be less than 0.9 and the 
polar angle of the missing momentum,
$\Theta_{\rm{miss}}$, must satisfy $|\cos \Theta_{\rm{miss}}| <$ 0.9.  

Four possible final states are considered: 
$\mathrm{hZ\ra  b\bar{b} \tau^+\tau^-}$, 
$\mathrm{hZ\ra  \tau^+\tau^- q\bar{q}}$, 
$\mathrm{hA\ra b\bar{b} \tau^+\tau^-}$ and 
$\mathrm{hA\ra \tau^+\tau^- b\bar{b}}$.
Each event is uniquely assigned to one channel
using mass and b-tag information, and  
final discriminants specific to each channel are 
constructed.
For the $\mathrm{hZ\ra b\bar{b} \tau^+\tau^-}$, 
$\mathrm{hA\ra b\bar{b} \tau^+\tau^-}$ and 
$\mathrm{hA\ra \tau^+\tau^- b\bar{b}}$ final 
states, the final discriminant is constructed from 
$m_{\rm{q\bar{q}}}$, $m_{\rm{\tau^+\tau^-}}$
and the b-tag variables of the two hadronic jets. 
For the $\mathrm{hZ\ra \tau^+\tau^- q\bar{q}}$ final state,
$m_{\rm{\tau^+\tau^-}}$ is used as the final discriminant.

\begin{figure}
\begin{tabular}{cc}
\hspace{-7mm}
\includegraphics*[width=0.5\textwidth]{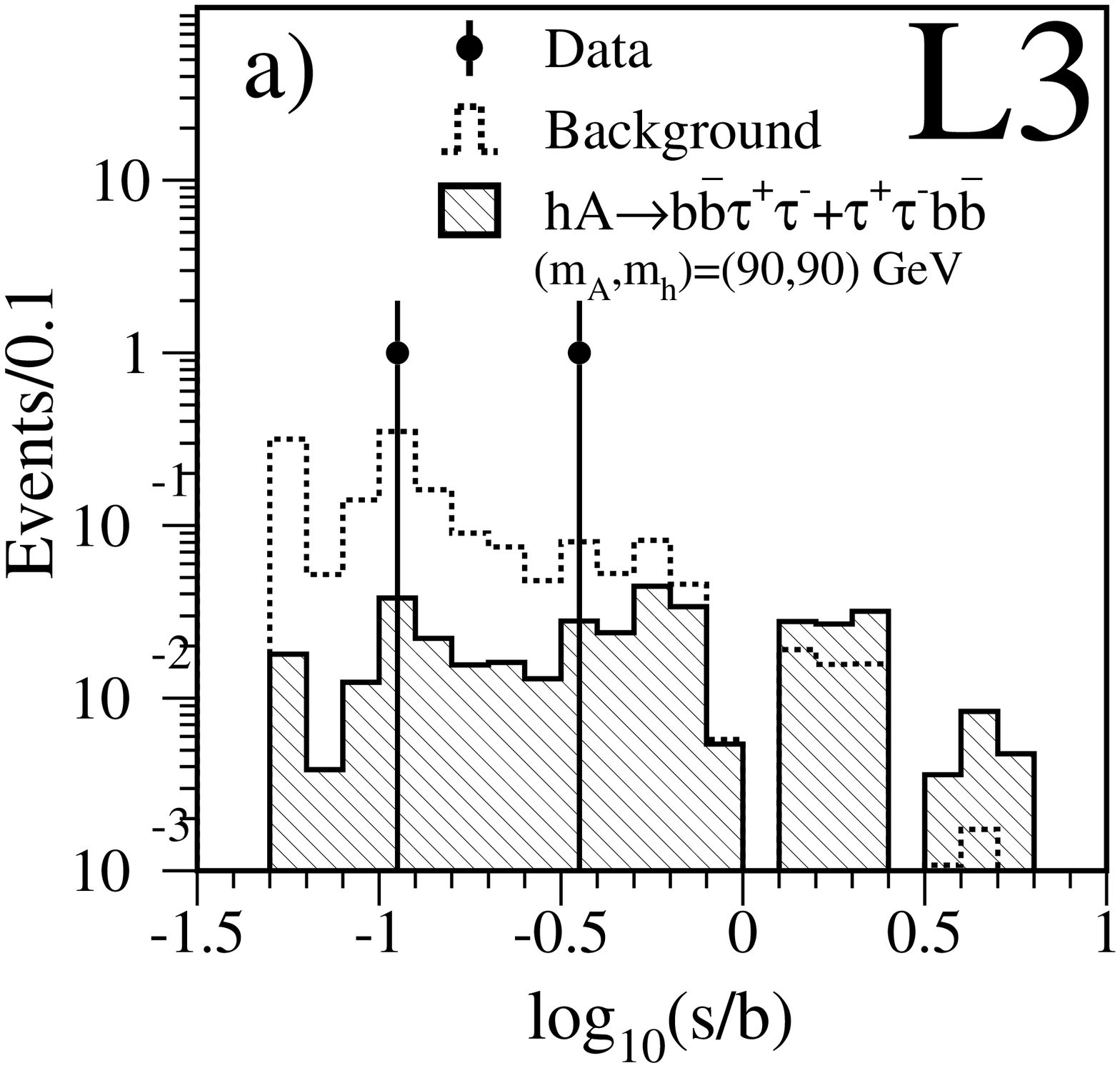} &
\hspace{-0mm}
\includegraphics*[width=0.5\textwidth]{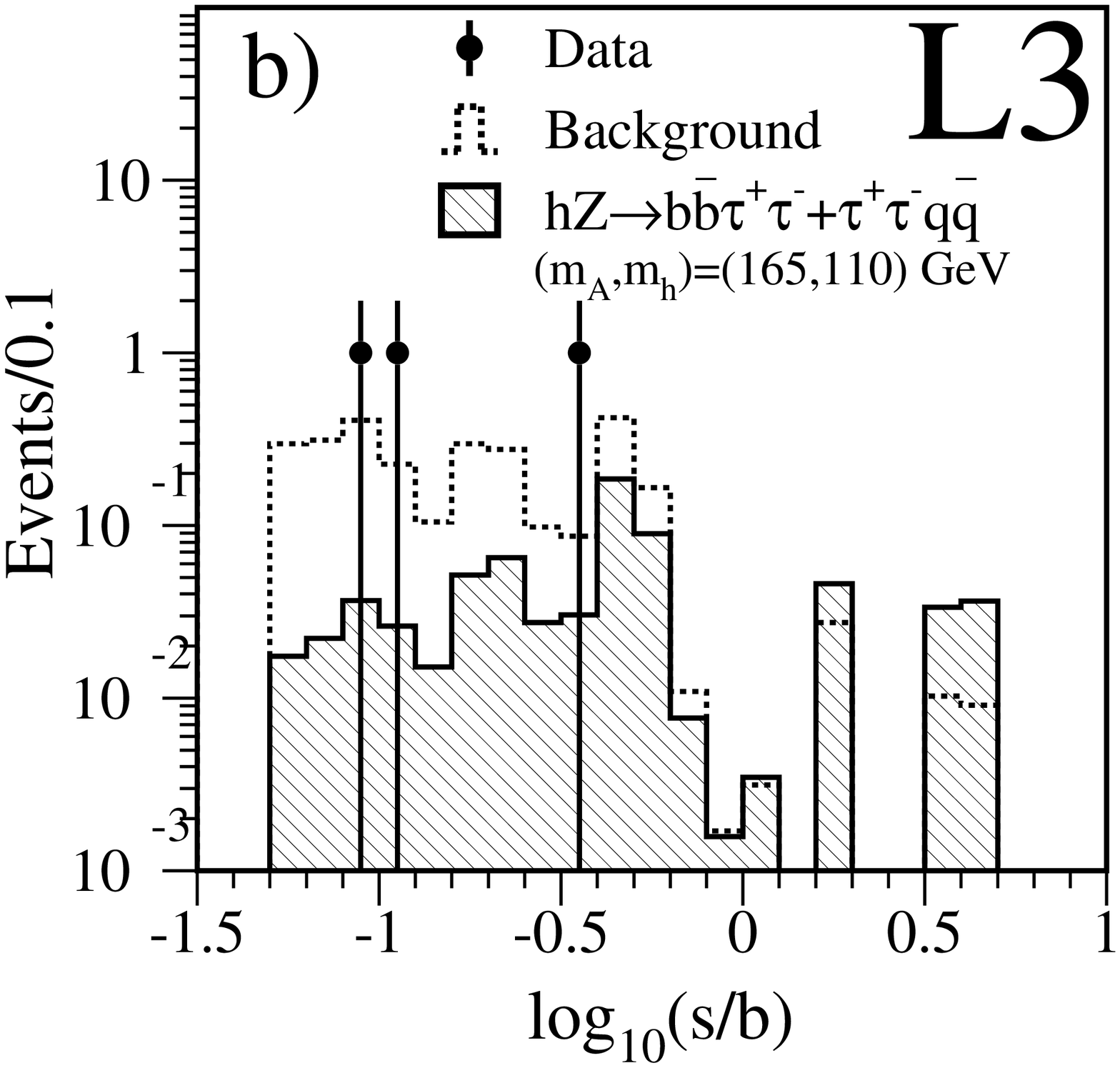} \\
\hspace{-7mm}
\includegraphics*[width=0.5\textwidth]{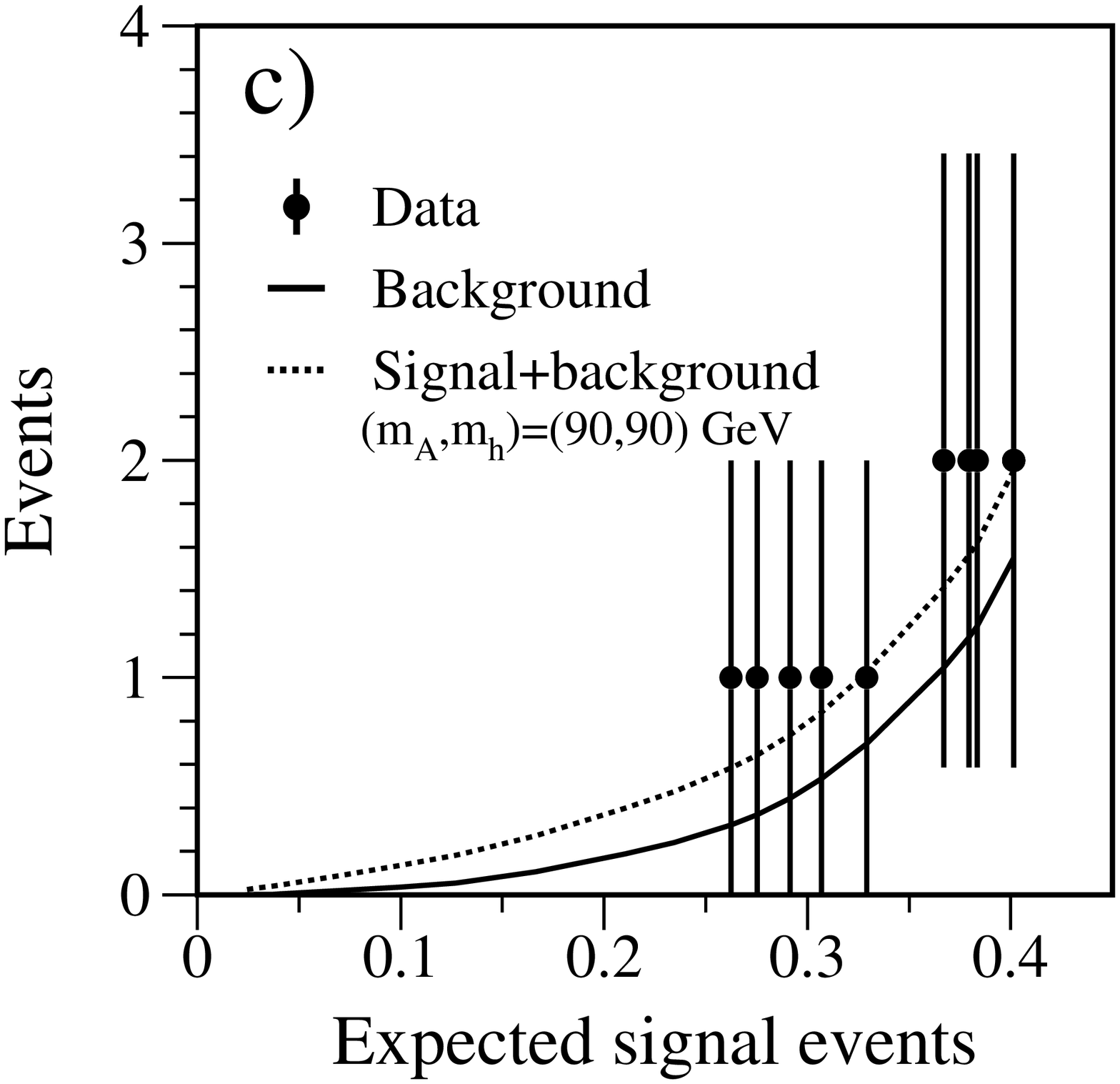} &
\hspace{-0mm}
\includegraphics*[width=0.5\textwidth]{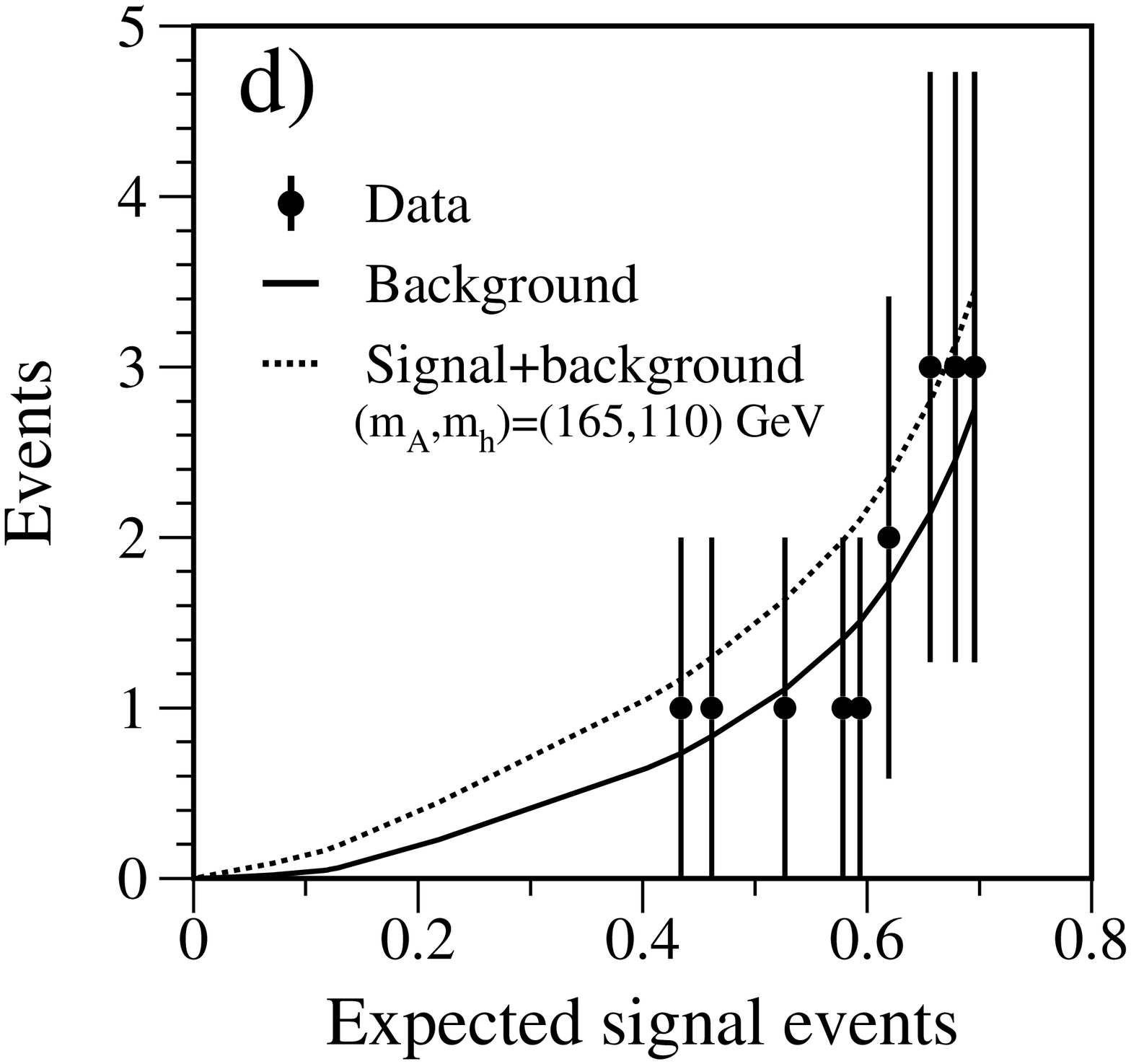} \\
\end{tabular}
\caption{
The distribution of data, expected 
background and expected signal events as a function of 
the logarithm of the signal-to-background ratio,
$\mathrm{log_{10}(s/b)}$, in the channels containing 
tau leptons for the Higgs boson mass 
hypotheses a) ($\mA$,$\mh$) = (90,90)\GeV and 
b) ($\mA$,$\mh$) = (165,110)\GeV.
Integrated distributions of data and expected background events as 
a function of the expected signal are shown in  
c) and d).
}
\label{fig:bbtt}
\end{figure}

Table~\ref{tab:bbtt} reports
the number of data, expected background and expected signal
events for the same Higgs boson masses 
chosen in the previous section.
For these hypotheses the distribution of the final discriminant
in terms of the signal-to-background ratio 
is shown in Figure~\ref{fig:bbtt}.  
Good agreement between the data and the expected background 
is found.

%
\subsection{The \boldmath{$\rm{hZ\ra AAq\bar{q}}$} Channel}
\begin{figure}
\begin{center}
\begin{tabular}{cc}
\hspace{-7mm}
\includegraphics*[width=0.5\textwidth]{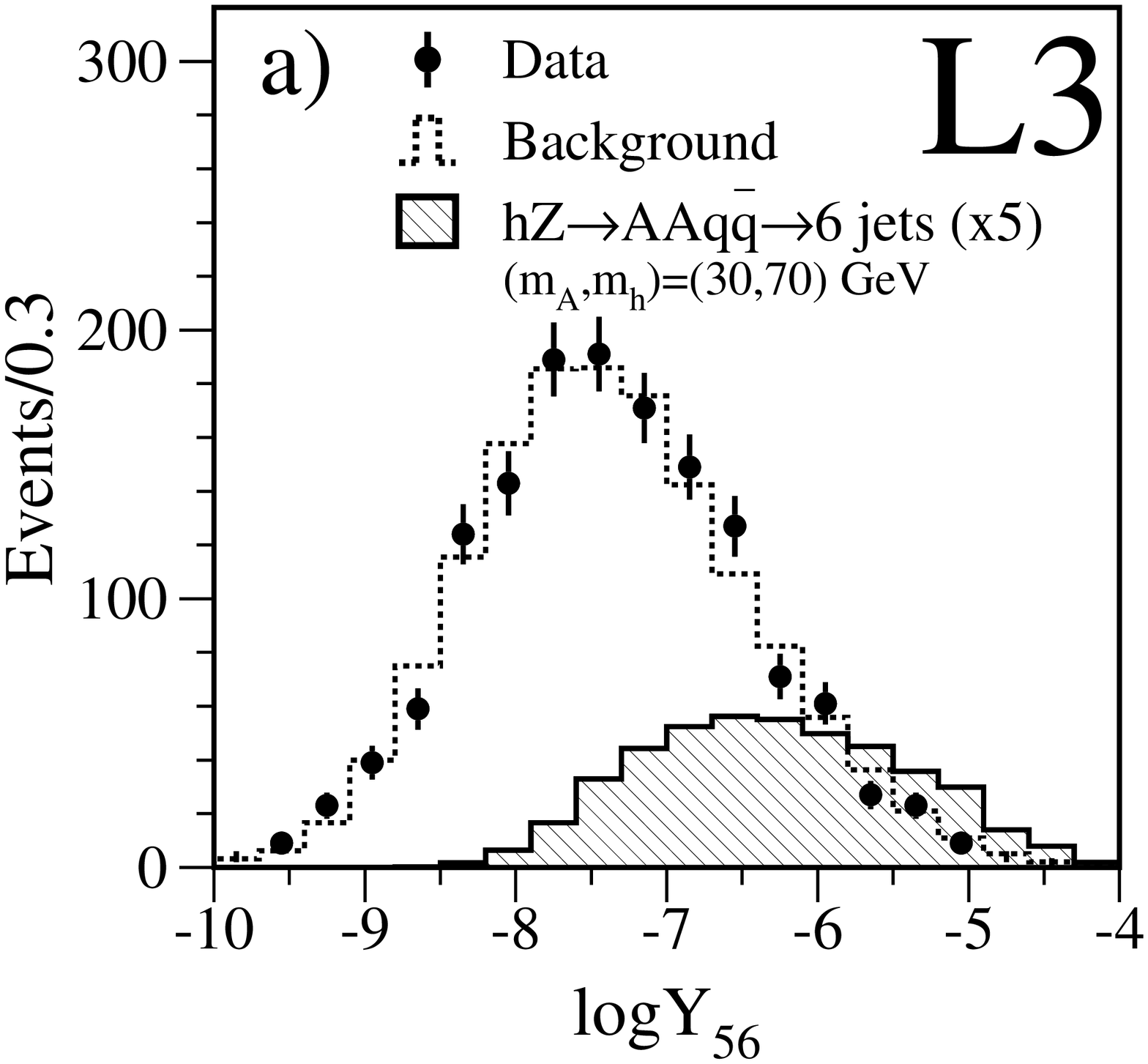} &
\hspace{-0mm}
\includegraphics*[width=0.5\textwidth]{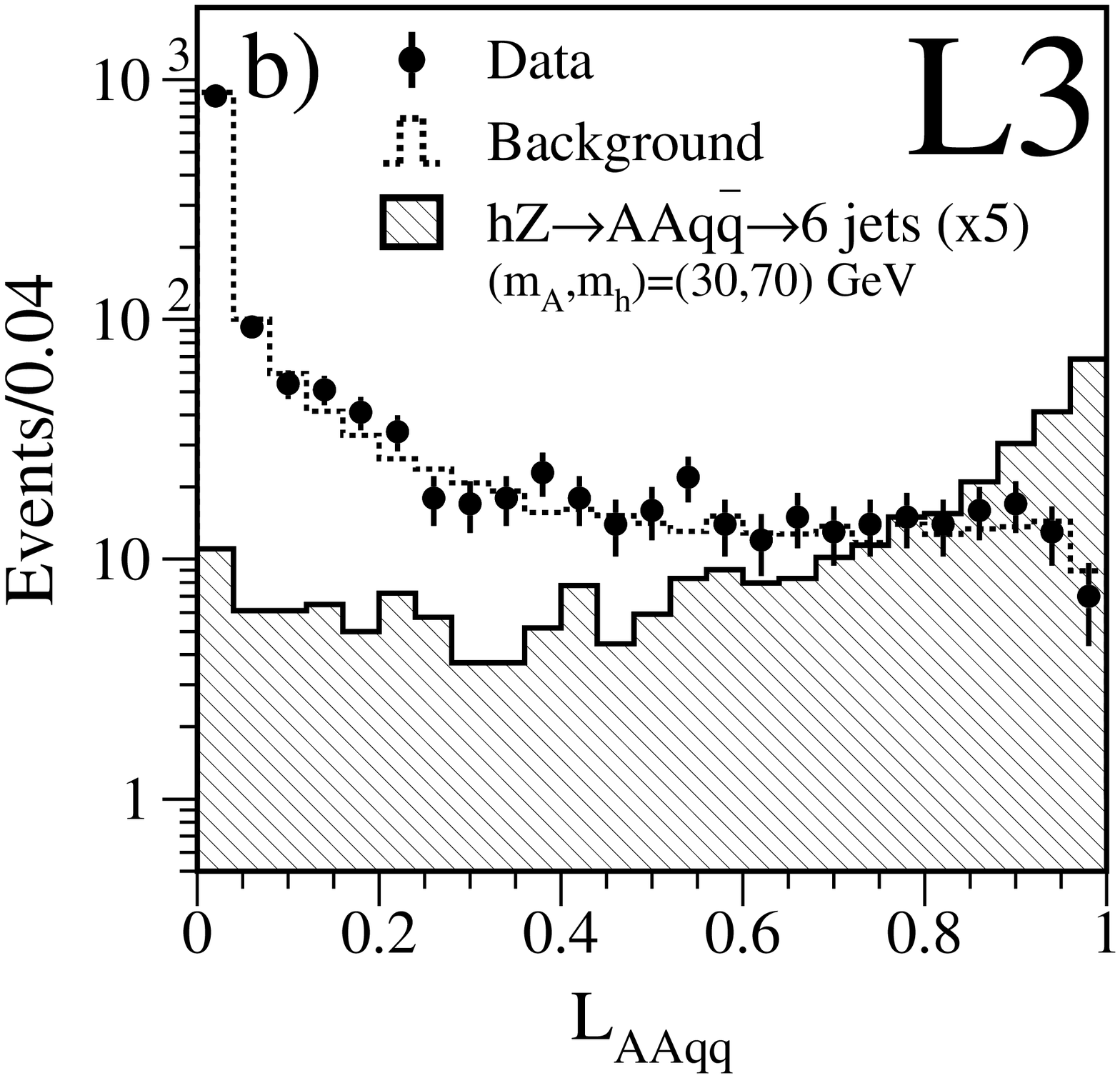} \\
\end{tabular}
\end{center}
\caption{Distributions 
of a) $\mathrm{logY_{56}}$  and b) $\mathrm{L_{AAqq}}$  
in the $\mathrm{hZ\ra AAq\bar{q}\ra q\bar{q}q^\prime\bar{q}^\prime
q^{\prime\prime}\bar{q}^{\prime\prime}}$ channel.
The points are data collected at $\sqrt{s} = 203-209\GeV$,
the dashed lines are   
the expected background and the hatched histograms are the
signal corresponding to the Higgs boson mass hypothesis ($\mA$,$\mh$) 
= (30,70)\GeV. The signal expectation 
is calculated within the ``no mixing'' 
scenario at \tanb = 0.75 and is multiplied by a factor of 5.}
\label{fig:zaa_plots}
\end{figure}

To improve the search sensitivity in the region of 
low \tanb and low $\mA$ where the
$\mathrm{h\ra AA}$ decay becomes dominant and the
$\mathrm{A\ra c\bar{c}}$ decay
replaces $\mathrm{A\ra b\bar{b}}$, 
a dedicated analysis is devised and performed 
on the data collected at $\mathrm{\sqrt{s}}$ = $189 - 209$ GeV.
This analysis aims to
select $\mathrm{hZ\ra AAq\bar{q}\ra q\bar{q}q^\prime\bar{q}^\prime
q^{\prime\prime}\bar{q}^{\prime\prime}}$ final states
and is derived from the analysis used in the four-jet 
channel. At the first stage, the  
same preselection as in the four-jet channel is applied 
with an additional cut on the event thrust, T $<$ 0.9.  
In the next step, a signal likelihood $\mathrm{L_{AAqq}}$
is built to distinguish the $\mathrm{hZ\ra AAq\bar{q}}$ 
signal from the $\mathrm{q\bar{q}(\gamma)}$ final states and
hadronic decays of W-pairs and Z-pairs.
This likelihood is constructed from the 
variables: $\mathrm{N_{trk}}$, 
$\mathrm{\gamma_{triple}}$, $\mathrm{logY_{34}}$,
the event sphericity, the absolute value of the cosine 
of the polar production angle, assuming the production of 
a pair of gauge bosons~\cite{l3_sm_paper} 
and $\mathrm{logY_{56}}$, where $\mathrm{Y_{56}}$ 
is the jet resolution parameter for which the 
event topology changes from five to six jets. 
Among these variables $\mathrm{N_{trk}}$ and
$\mathrm{logY_{56}}$ have the most discriminating 
power between the $\mathrm{hZ\ra AAq\bar{q}}$
signal and four-fermion and two-fermion backgrounds.
The likelihood $\mathrm{L_{AAqq}}$ is used 
as the final discriminant.  
No evidence for the $\mathrm{hZ\ra AA q\bar{q}}$ signal is found in data.
As an example, Figure~\ref{fig:zaa_plots} shows the distributions 
of $\mathrm{logY_{56}}$ and $\mathrm{L_{AAq\bar{q}}}$
for data, the
expected background and the signal corresponding to the
Higgs boson mass hypothesis ($\mA$,$\mh$) = (30,70) GeV.
    
%
\section{Results}

The analyses presented in this Letter are combined with 
the $\mathrm{hZ\ra b\bar{b}\nu\bar{\nu}}$ and 
$\mathrm{hZ\ra b\bar{b}\ell^+\ell^-}$ $\mathrm{(\ell=e,\mu)}$ analyses used 
in the Standard Model Higgs searches~\cite{l3_sm_paper}. 
The results of previous searches~\cite{l3_2000,l3_1999} at lower 
$\mathrm{\sqrt{s}}$ are also included. 
The final discriminant distributions 
obtained in each search channel at each 
centre-of-mass energy are used to evaluate the presence of 
a signal in the data. No evidence for a signal is found and 
the search results are interpreted in terms of an exclusion of 
MSSM parameter regions. The statistical 
procedure adopted for the interpretation of the data 
and the definition of the confidence level $\mathrm{CL_{s}}$ 
are described in Reference~\citen{ratio_method}.  
The analysis performance is quantified with the 
expected median confidence level, $\mathrm{CL_{med}}$, which 
is obtained from $\mathrm{CL_{s}}$ by replacing the observed 
value of the test-statistic $-2\rm{lnQ}$ by its background 
median value.

Systematic and statistical uncertainties on the signal
and on the background are incorporated in the confidence
level calculations as described in Reference~\citen{systematics}.
The main sources of systematic uncertainties  are the detector 
resolution, the selection procedure, theoretical uncertainties
and Monte Carlo statistics. The overall systematic 
uncertainty is estimated to range from 3\% to 6\% on the 
expected signal, and from 7\% to 15\% 
on the background, depending on the search channel.

\subsection{Limits in the ``\boldmath{$\mh-$}max'' and ``No mixing'' Scenarios}
\begin{figure}
\begin{tabular}{cc}
\hspace{-7mm}
\includegraphics*[width=0.5\textwidth]{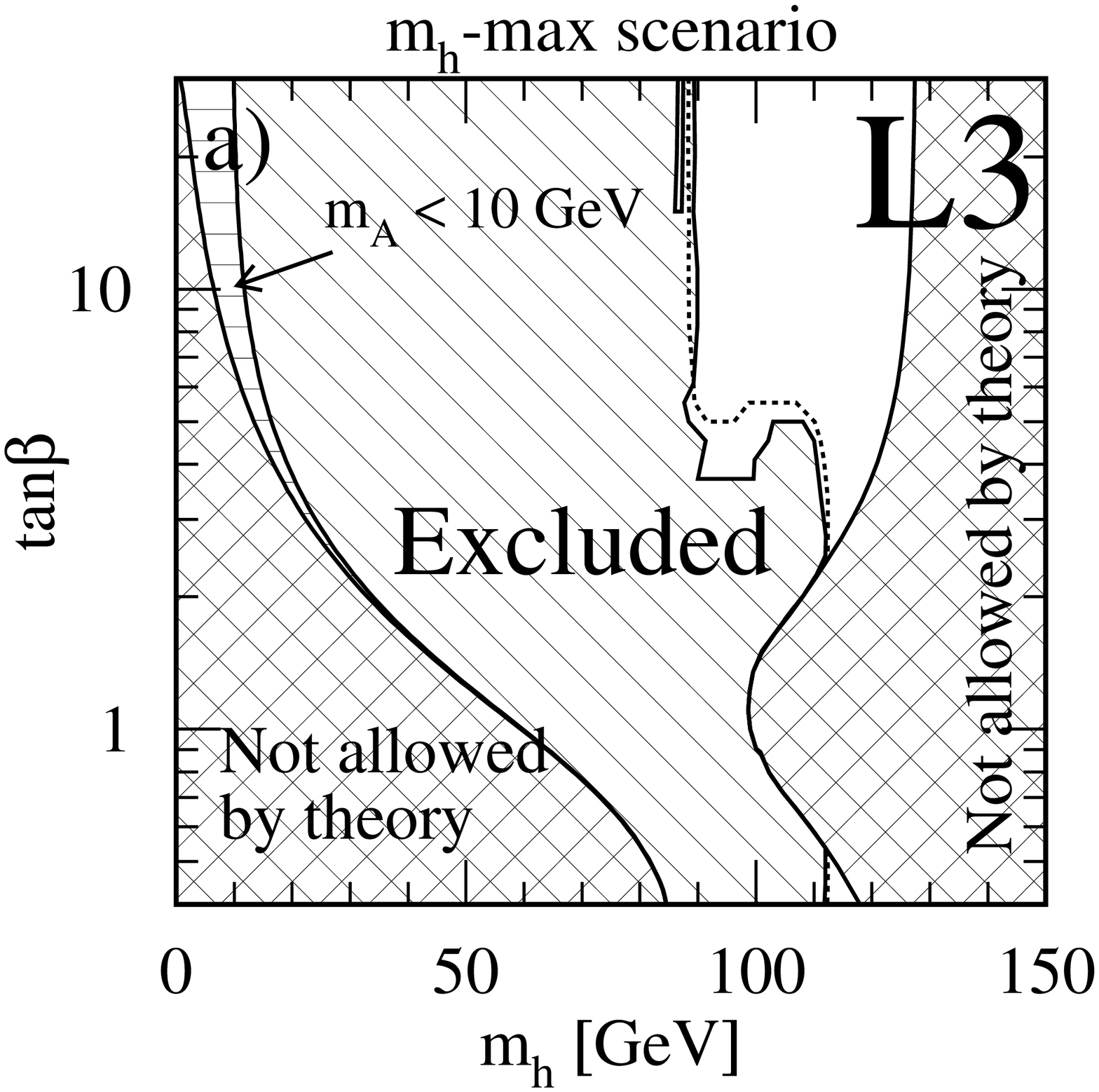} &
\hspace{-0mm}
\includegraphics*[width=0.5\textwidth]{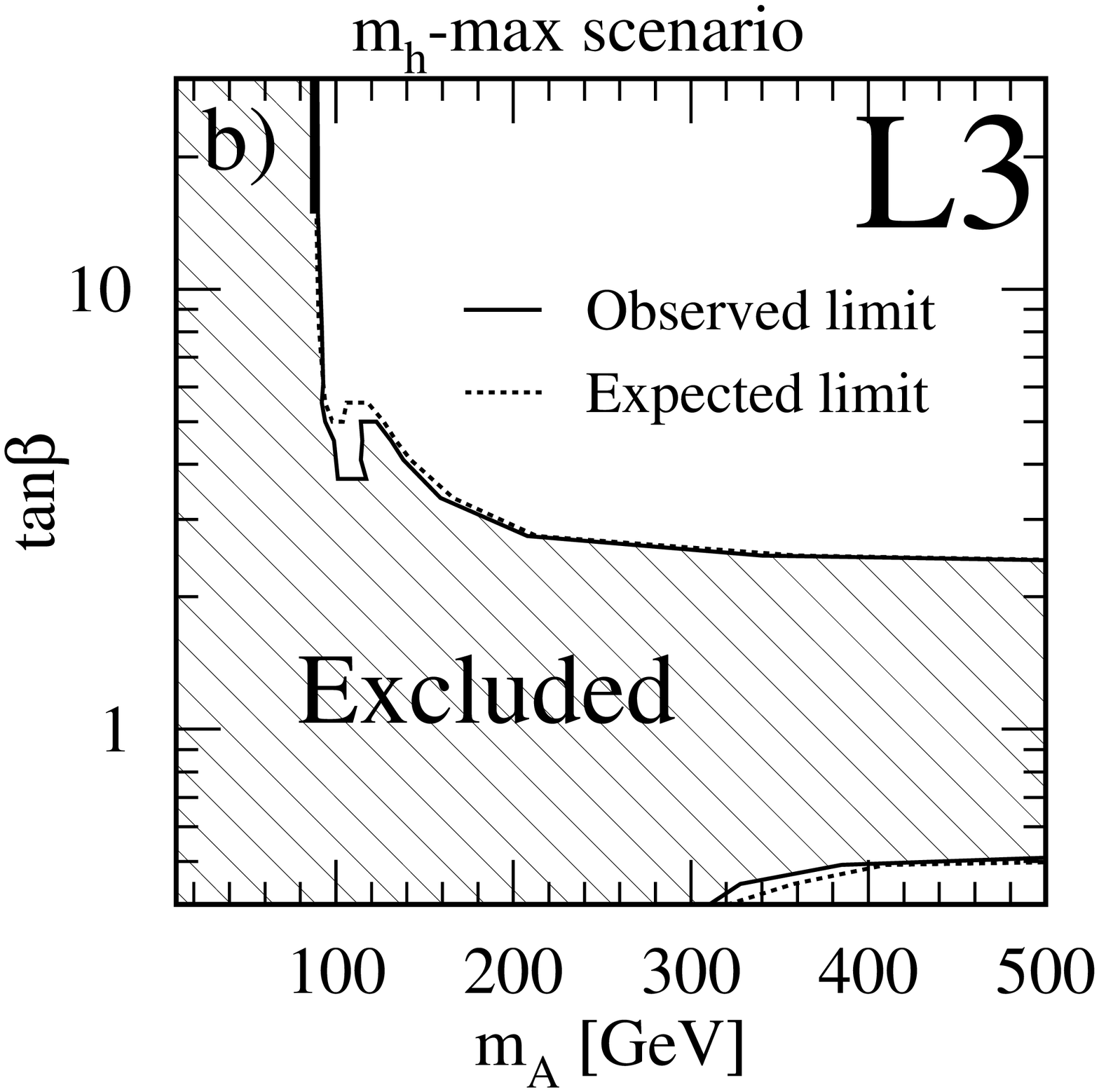} \\
\hspace{-7mm}
\includegraphics*[width=0.5\textwidth]{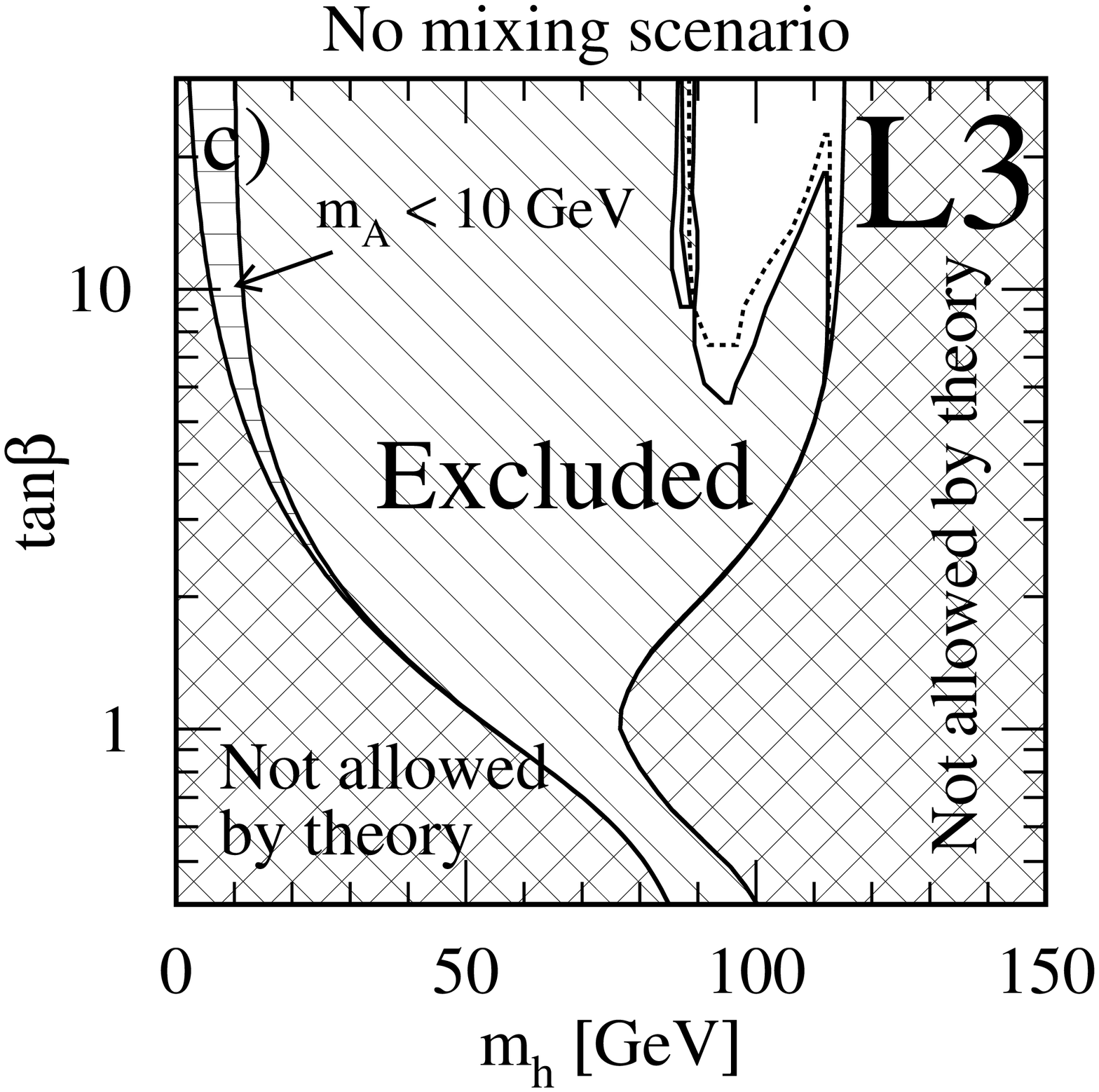} &
\hspace{-0mm}
\includegraphics*[width=0.5\textwidth]{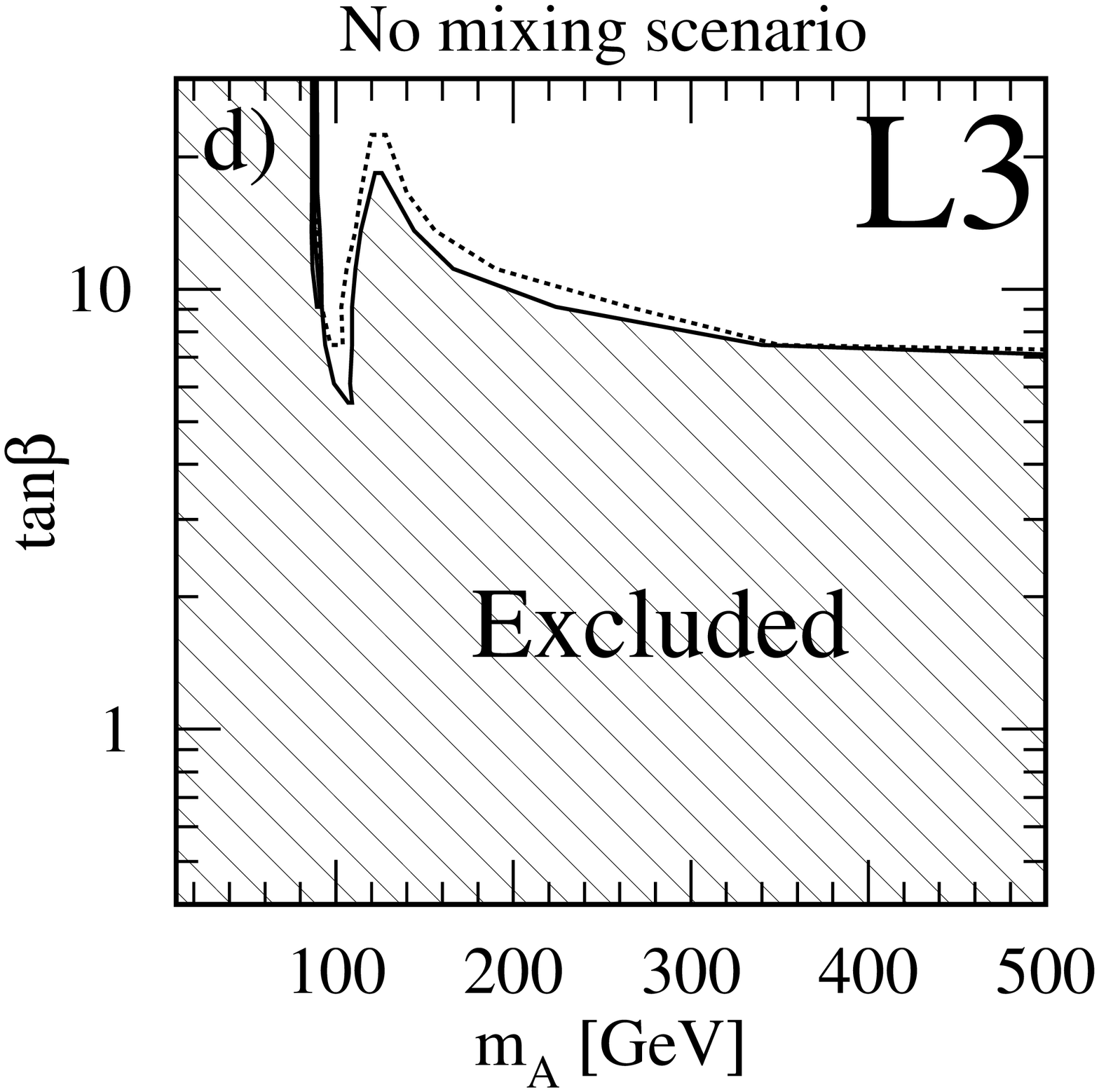} \\
\end{tabular}
    \caption[]{\label{fig:minmax}
              Exclusion contours in the (\tanb,$\mh$) 
              and (\tanb,$\mA$) planes at 95\% confidence level for 
              the ``$\mh-$max'' and ``no mixing''
              scenarios. The hatched area represents the 
              exclusion and the crossed area is not allowed by 
              theory. The horizontally hatched area corresponds to 
              $\mA<10$ GeV and was previously excluded 
              by LEP~\cite{ma30_opal}. The dashed line indicates 
              the expected exclusion in the absence of a signal.
            }

\end{figure}

Figure~\ref{fig:minmax} shows the
area of the (\tanb,$\mh$) 
and (\tanb,$\mA$) planes excluded at 95\% confidence level 
for the ``$\mh-$max'' and ``no mixing'' scenarios.
In  the ``$\mh-$max'' scenario, lower limits
on the masses of the h and A bosons are set at 95\% confidence level as:
\begin{displaymath}
 \mA > 86.5\GeV,\hspace{10mm}\mh > 86.0\GeV,
\end{displaymath}
for every \tanb value considered. 
The expected values 
in the absence of a signal are 
$\mA > 88.6\GeV$ and $\mh > 88.4\GeV$.
For 0.55 $<$ \tanb $<$ 2.2 the A boson is
excluded up to a mass of 1 TeV
thus allowing to rule out this \tanb 
range.

In the ``no mixing'' scenario, the combined results 
establish lower mass bounds at 95\% confidence level of:
\begin{displaymath}
\mA > 86.3\GeV,\hspace{10mm}\mh > 85.5\GeV.
\end{displaymath}
The expected limits are
$
\mA > 88.6\GeV$ and $\mh > 88.5\GeV.
$
Here the \tanb range between 0.4 and 5.4 is excluded at 95\%
confidence level, also for $\mA$ up to 1 TeV. 
A downward fluctuation of about $1\sigma$ compared
to the background expectation is observed in the data at
($\mA$,$\mh$) $\sim$ (90,90) GeV. 
There is a deficit of candidates 
in the $\mathrm{hA\ra b\bar{b}b\bar{b}}$ channel
at $\mathrm{\sqrt{s}}$ = $203-209$ GeV resulting 
in an observation of a $\mathrm{CL_s}$ value smaller than 
5\% although no exclusion of this region at 95\% confidence level is expected. 
This effect explains the irregularity in the exclusion plots
at high \tanb and $\mh$ between 86 GeV and 91 GeV.
An excess in the $\mathrm{hZ\ra b\bar{b}\ell^+\ell^-(\ell=e,\mu)}$
channel at $\mh\sim 90$ GeV and in the $\mathrm{hZ\ra b\bar{b}q\bar{q}}$
channel in the $\mh$ range between 90 GeV and 95 GeV, from 
the $\mathrm{\sqrt{s}=192-202}$ GeV data, 
result in a sizable reduction of the
excluded range of \tanb for the mass range
90 GeV $\lesssim$ $\mh$ $\lesssim$ 100 GeV in both 
scenarios. The area at \tanb $<$ 0.8 and 
$\mA$ $<$ 40 GeV which previously was not excluded 
in the ``no mixing'' scenario~\cite{l3_2000} is now excluded 
using the results of  
the $\mathrm{hZ\ra AAq\bar{q}}$ analysis\footnote{The 
$\mathrm{hZ\ra AAq\bar{q}}$ analysis is used instead 
of the four-jet one whenever it provides better sensitivity, 
{\it{i.e.}} gives smaller values of $\mathrm{CL_{med}}$.}.

\subsection{Limits in the ``Large$-\mu$'' Scenario}
\begin{figure}
\begin{tabular}{cc}
\hspace{-7mm}
\includegraphics*[width=0.5\textwidth]{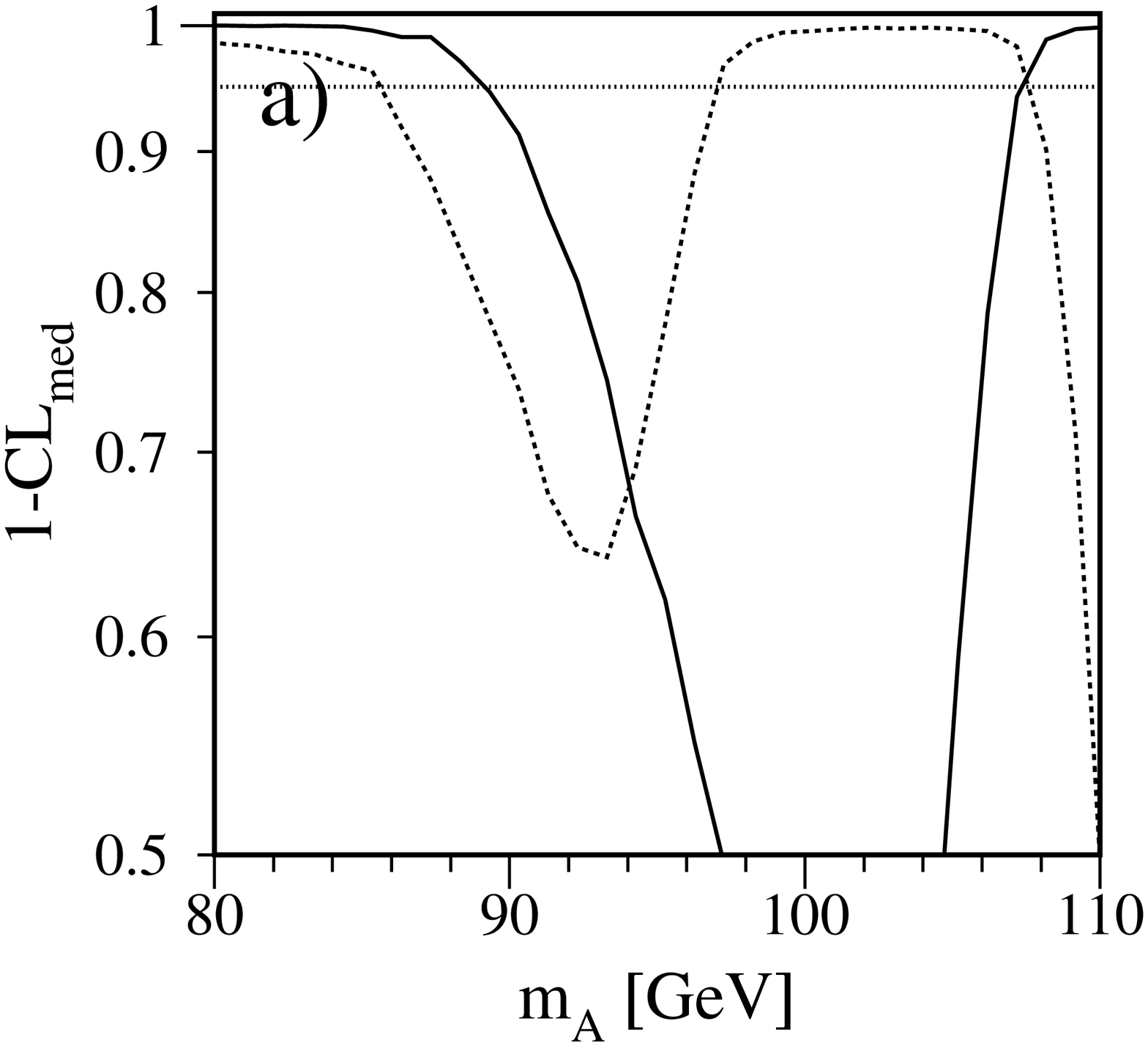} &
\hspace{-0mm}
\includegraphics*[width=0.5\textwidth]{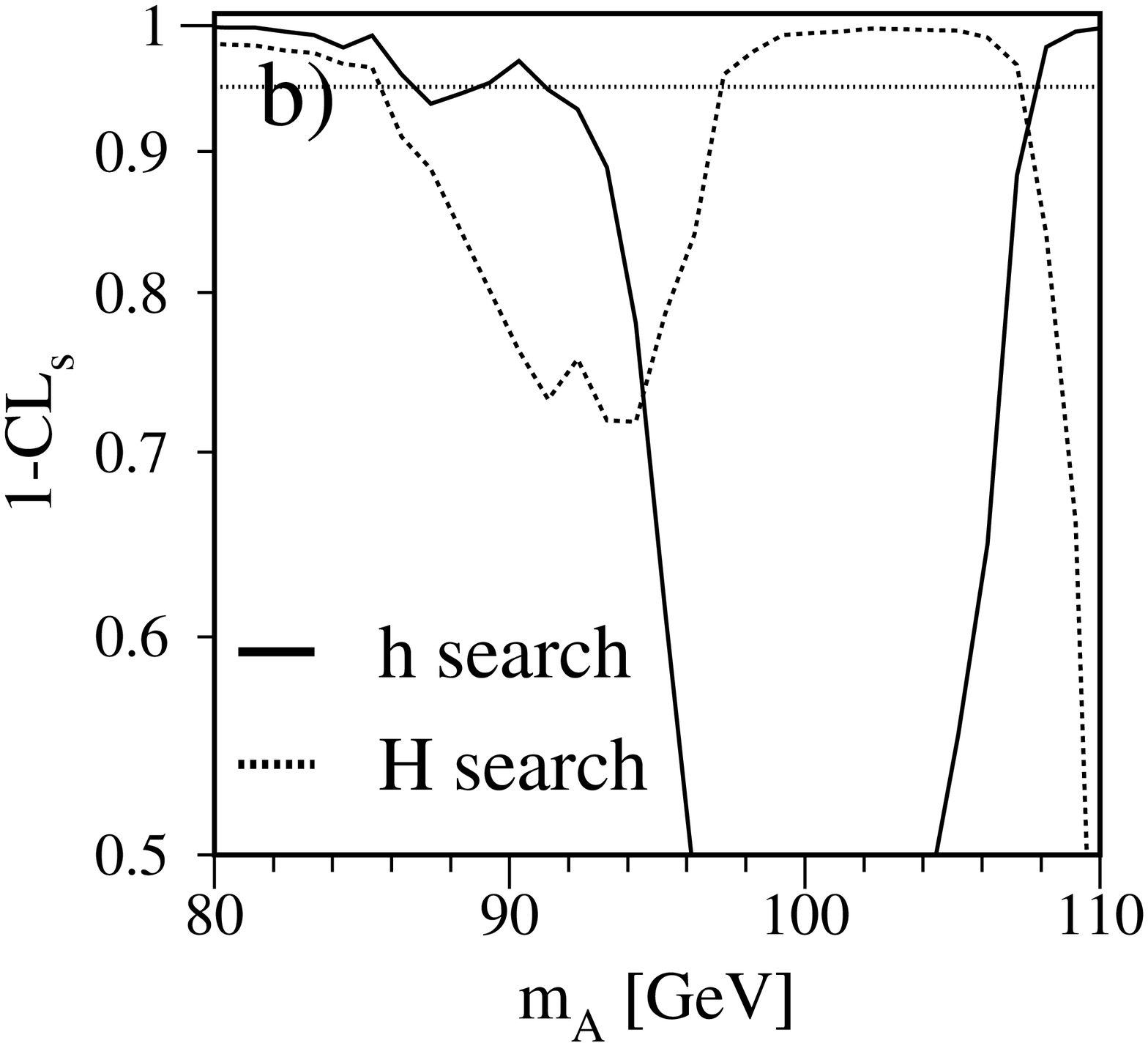} \\
\hspace{-7mm}
\includegraphics*[width=0.5\textwidth]{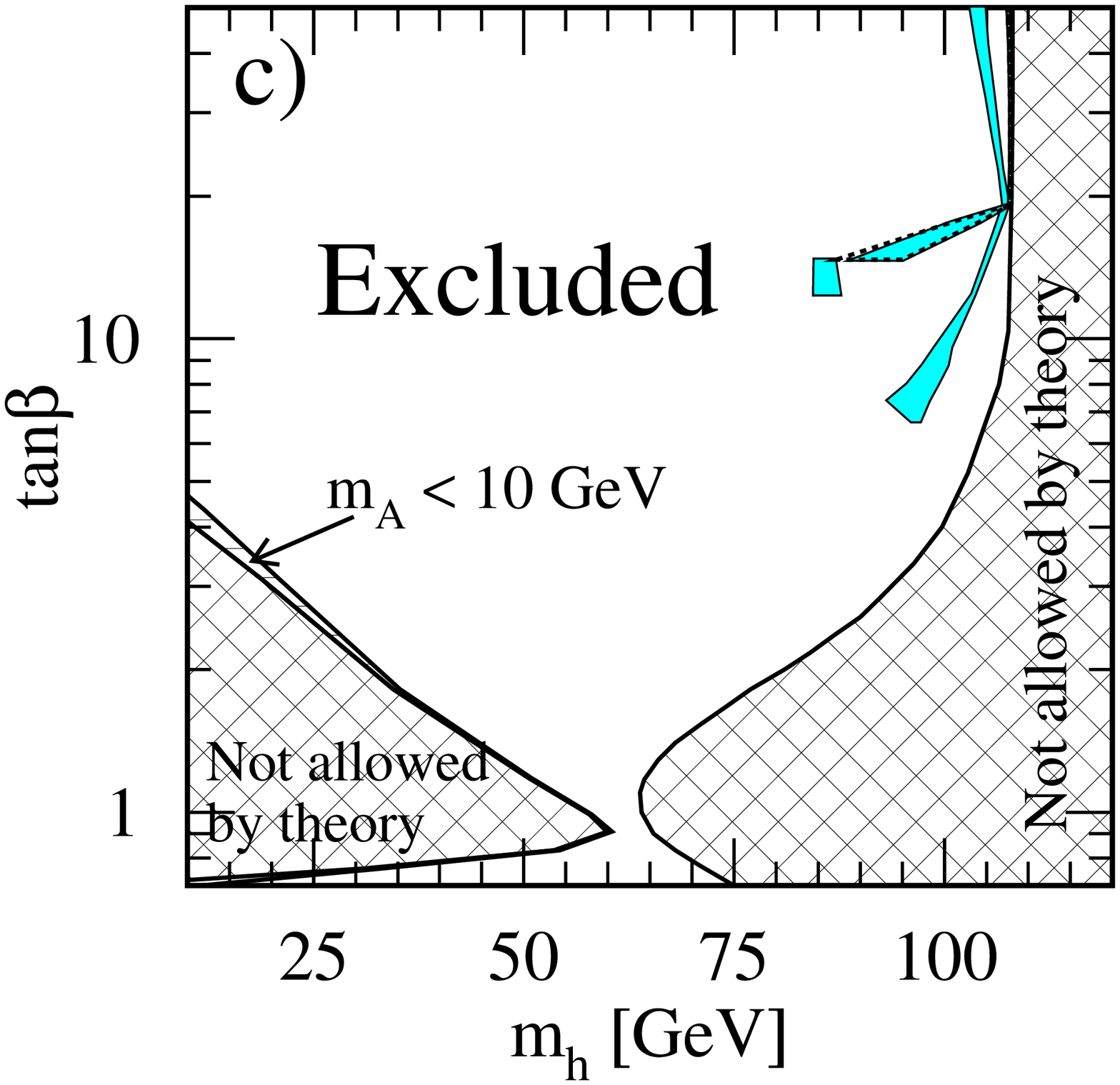} &
\hspace{-0mm}
\includegraphics*[width=0.5\textwidth]{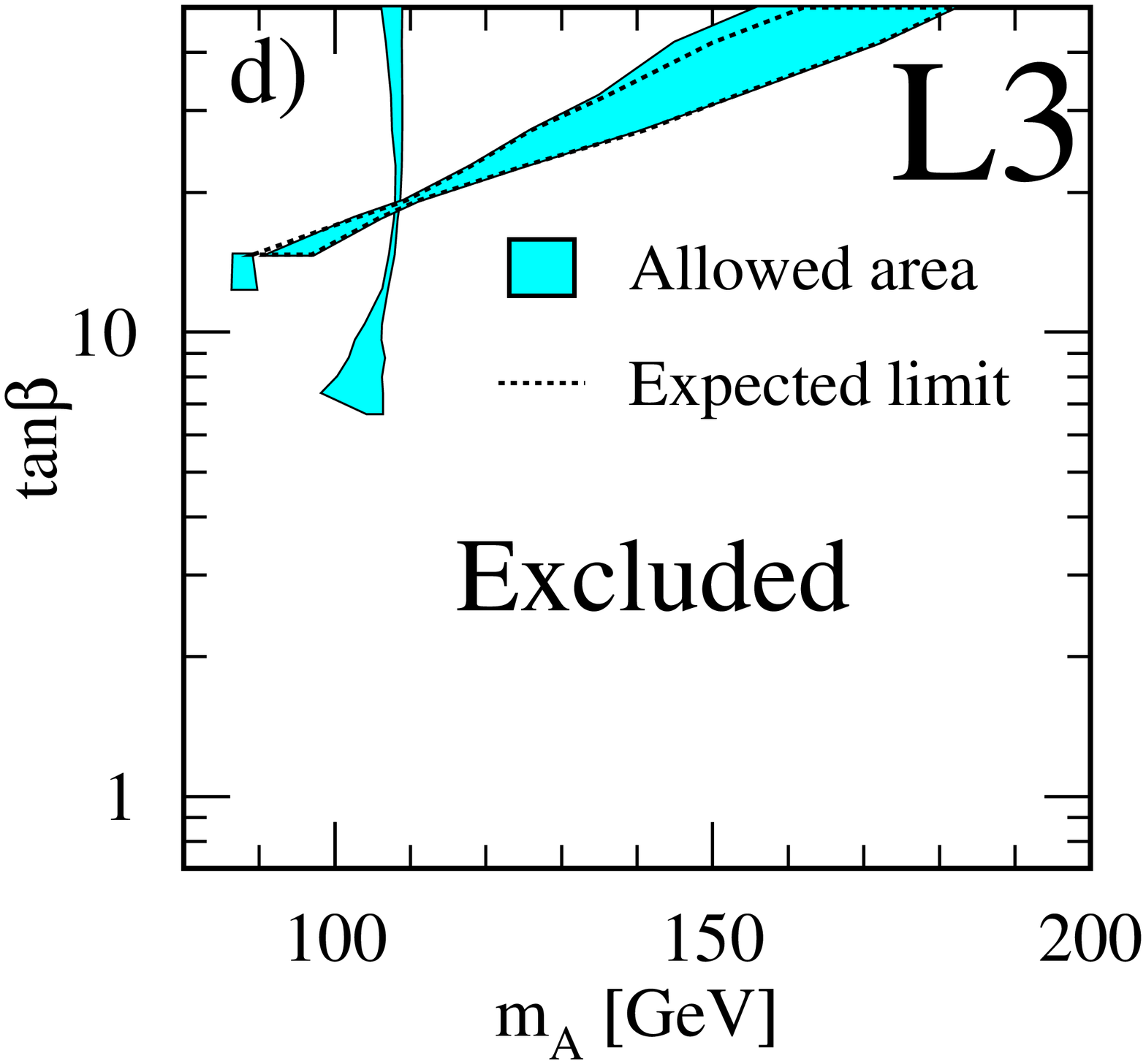} \\
\end{tabular}
\caption{ Confidence levels a) ($1-\mathrm{CL_{med}}$) and b)
          ($1-\mathrm{CL_{s}}$) as a function of $\mA$ 
          at $\tanb=15$ obtained for
          the h boson (solid line), 
          and the H boson (dashed line) searches in the 
          ``large$-\mu$'' scenario.
          Exclusion contours in the c) (\tanb,$\mh$)  and 
          d) (\tanb,$\mA$) planes for the ``large$-\mu$'' scenario.
          The crossed area is theoretically inaccessible,
          the open area is excluded at 95\% confidence level and
          the shaded area is experimentally allowed.
          The dashed line represents the expected boundary of 
          the allowed region in the absence of a signal.}
\label{fig:lmu}
\end{figure}

In the ``large$-\mu$'' scenario there are
regions of the (\tanb,$\mA$) plane
where the hA process is inaccessible and the hZ process 
is suppressed by a small value of $\sin^2(\beta-\alpha)$. 
However, the heavy CP even Higgs boson H is expected
to be produced there via the 
Higgs-strahlung process. Hence, the loss of the sensitivity 
for the h boson can be compensated by reinterpreting the 
hZ analyses in the context of the HZ search.
This is illustrated in Figure~\ref{fig:lmu}a 
which presents the ($1-\mathrm{CL_{med}}$) confidence level
calculated as a function of $\mA$ at $\tanb=15$ in 
the context of the searches for the \h and \bigH bosons. 
Searches for the h boson alone lack sensitivity in
the $\mA$ range $89\GeV \lesssim \mA \lesssim 108\GeV$.
The inclusion of the HZ search results extends
the region of sensitivity leaving 
only the range $89\GeV \lesssim \mA \lesssim 97\GeV$ unexcluded. 
However, since all the analyses, except for 
the $\mathrm{hZ\ra \tau^+\tau^- q\bar{q}}$ channel, are
optimised for the Higgs boson decays into $\mathrm{b\bar{b}}$, 
they do not provide sufficient
sensitivity to the parameter regions where the
effective couplings $\mathrm{Hb\bar{b}}$ and 
$\mathrm{hb\bar{b}}$ are suppressed~\cite{qcd_susy_1}.
The observed exclusion is presented in Figure~\ref{fig:lmu}b.

Exclusion plots in the (\tanb,$\mh$) and (\tanb,$\mA$) planes 
for the ``large$-\mu$'' scenario are presented in Figures~\ref{fig:lmu}c and 
~\ref{fig:lmu}d, respectively.  
Limits on Higgs boson masses
are derived as:
\begin{displaymath}
\mh > 84.5\GeV,\hspace{10mm}\mA > 86.5\GeV.
\end{displaymath}
The expected values are 87.2\GeV and 89.2\GeV, respectively.
Furthermore, the range $0.7<\tanb<6.7$ is excluded, for values of 
$\mA$ up to 400\GeV.
The allowed area between $\tanb=15$ and 
$\tanb=50$ corresponds to  
reduced $\mathrm{Hb\bar{b}}$ or $\mathrm{hb\bar{b}}$ 
couplings. The unexcluded region at $\mA$ $\sim$ 88 GeV and $\tanb\sim15$
is caused by a slight upward fluctuation in the data,  
coming mainly from $\mathrm{hA\ra b\bar{b}b\bar{b}}$ 
candidates selected in the data at $\mathrm{\sqrt{s}>203}$ GeV. 
The allowed vertical narrow band 
at $\mA$ = $107 - 110$ GeV  
and \tanb $\gtrsim$ 10 represents the region where
the hA production is kinematically inaccessible and   
$\mathrm{\cos^2(\beta-\alpha)}$ $\approx$ $\mathrm{\sin^2(\beta-\alpha)}$
$\approx$ 0.5 so that both the hZ and HZ production cross sections
are reduced by a factor of 2 with respect to $\mathrm{\sigma_{HZ}^{SM}}$. 
Although the L3 combined search has a sensitivity for exclusion 
of this critical region, as can be seen from 
Figure~\ref{fig:lmu}a, the expected median confidence 
level is only slightly lower than 5\% and an 
insignificant upward fluctuation observed in the data 
pushes the observed confidence level, $\mathrm{CL_s}$, 
above 5\%, thus not allowing to exclude this region at 95\% confidence level. 
Finally, the allowed area at \tanb between 6.7 and 10 and 
$\mh$ between 90 and 100 GeV arises due to the excesses
already discussed.

In conclusion, no evidence for neutral Higgs bosons of the MSSM is
found and large regions of its parameter space are excluded.

\bibliographystyle{/l3/paper/biblio/l3stylem}

\begin{mcbibliography}{10}

\bibitem{mssm_1}
H. P. Nilles,
\newblock  Phys. Rep. {\bf 110}  (1984) 1;\relax
H. E. Haber and G. L. Kane,
\newblock  Phys. Rep. {\bf 117}  (1985) 75;\relax
R. Barbieri,
\newblock  Riv. Nuovo Cim. {\bf 11 n$^\circ$4}  (1988) 1\relax
\relax
\bibitem{higgs_hunters}
J.F. Gunion \etal,
\newblock  in The Higgs Hunter's Guide,  (Addison Wesley, 1990), p. 191\relax
\relax
\bibitem{top_mass}
D.E. Groom \etal,
\newblock  Eur. Phys. J. {\bf C 15}  (2000) 1\relax
\relax
\bibitem{l3_2000}
L3 Collaboration, M. Acciarri \etal,
\newblock  Phys. Lett. {\bf B 503}  (2001) 21\relax
\relax
\bibitem{l3_1999}
L3 Collaboration, M. Acciarri \etal,
\newblock  Phys. Lett. {\bf B 471}  (1999) 321\relax
\relax
\bibitem{opal_mssm}
OPAL Collaboration, G. Abbiendi \etal,
\newblock  Eur. Phys. J. {\bf C 12}  (2000) 567;\relax
ALEPH Collaboration, A. Heister \etal,
\newblock  Phys. Lett. {\bf B 526}  (2002) 191;\relax
DELPHI Collaboration, J. Abdallah \etal,
\newblock  Eur. Phys. J. {\bf C 23}  (2002) 409\relax
\relax
\bibitem{l3_det}
L3 Collab., B. Adeva \etal, Nucl. Instr. Meth. {\bf A 289} (1990) 35;\\ J.A.
  Bakken \etal, Nucl. Instr. Meth. {\bf A 275} (1989) 81;\\ O. Adriani \etal,
  Nucl. Instr. Meth. {\bf A 302} (1991) 53;\\ B. Adeva \etal, Nucl. Instr.
  Meth. {\bf A 323} (1992) 109;\\ K. Deiters \etal, Nucl. Instr. Meth. {\bf A
  323} (1992) 162;\\ M. Chemarin \etal, Nucl. Instr. Meth. {\bf A 349} (1994)
  345;\\ M. Acciarri \etal, Nucl. Instr. Meth. {\bf A 351} (1994) 300;\\ G.
  Basti \etal, Nucl. Instr. Meth. {\bf A 374} (1996) 293;\\ A. Adam \etal,
  Nucl. Instr. Meth. {\bf A 383} (1996) 342\relax
\relax
\bibitem{mssm_benchmark}
M. Carena \etal,
\newblock  Preprint CERN-TH/99-374\relax
\relax
\bibitem{heinemeyer_1}
S.Heinemeyer, W.Hollik, G. Weiglein,
\newblock  Eur. Phys. J. {\bf C9}  (1999) 343;\relax
S. Heinemeyer, W. Hollik, G. Weiglein,
\newblock  Phys. Rev. {\bf D 58}  (1998) 091701;\relax
S. Heinemeyer, W. Hollik, G. Weiglein,
\newblock  Phys. Lett. {\bf B 440}  (1998) 296\relax
\relax
\bibitem{carena_1}
M. Carena, M. Quiros and C. E. M. Wagner,
\newblock  Nucl. Phys. {\bf B 461}  (1996) 407;\relax
H. Haber, R. Hempfling, A. Hoang,
\newblock  Z. Phys. {\bf C 75}  (1997) 539\relax
\relax
\bibitem{hzha}
P. Janot,
\newblock  ``The HZHA generator", in ``Physics at LEP2",
\newblock  CERN Report 96-01 (1996), Version 3, released in December 1999,\\
  {\tt{http://alephwww.cern.ch/\~\,janot/Generators.html}}\relax
\relax
\bibitem{pythia}
PYTHIA versions 5.722 and 6.1 are used.\\ T. Sj{\"o}strand, Preprint
  CERN-TH/7112/93 (1993), revised August 1995; {Comp. Phys. Comm.} {\bf 82}
  (1994) 74; Preprint hep-ph/0001032 (2000)\relax
\relax
\bibitem{kk2f}
KK2f version 4.13 is used; \\ S.~Jadach, B.F.L.~Ward and Z.~W\c{a}s, Comp.
  Phys. Comm. {\bf 130} (2000) 260\relax
\relax
\bibitem{koralw}
KORALW version 1.33 is used.\\ S. Jadach {\em et~al.}, Comp. Phys. Comm. {\bf
  94} (1996) 216;\\ S. Jadach {\em et~al.}, Phys. Lett. {\bf B 372} (1996)
  289\relax
\relax
\bibitem{koralz}
KORALZ version 4.02 is used. \\ S. Jadach, B.F.L. Ward and Z. W\c{a}s, {Comp.
  Phys. Comm.} {\bf 79} (1994) 503\relax
\relax
\bibitem{phojet}
PHOJET version 1.05 is used. \\ R.~Engel, Z. Phys. {\bf C 66} (1995) 203;\\
  R.~Engel and J.~Ranft, {Phys. Rev.} {\bf D 54} (1996) 4244\relax
\relax
\bibitem{excalibur}
F. A. Berends, R. Pittau and R. Kleiss,
\newblock  Comp. Phys. Comm. {\bf 85}  (1995) 437\relax
\relax
\bibitem{geant}
GEANT version 3.15 is used;
R. Brun \etal,
\newblock  Preprint CERN DD/EE/84-1 (1984), revised 1987\relax
\relax
\bibitem{gheisha}
H. Fesefeldt,
\newblock  Report RWTH Aachen PITHA 85/02 (1985)\relax
\relax
\bibitem{l3_sm_paper}
L3 Collaboration, P. Achard \etal,
\newblock  Phys. Lett. {\bf B 517}  (2001) 319\relax
\relax
\bibitem{DURHAM}
S. Bethke \etal,
\newblock  Nucl. Phys. {\bf B 370}  (1992) 310\relax
\relax
\bibitem{ratio_method}
A. Read, {{``Modified Frequentist Analysis of Search Results''}} in
  {\it{Workshop on Confidence Limits}}, eds. F. James, L.Lyons and Y.Perrin,
  CERN 2000-05, p. 81\relax
\relax
\bibitem{systematics}
R. D. Cousins and V. L. Highland,
\newblock  Nucl. Inst. Meth. {\bf A 320}  (1992) 331\relax
\relax
\bibitem{qcd_susy_1}
W. Loinaz and J.D. Wells,
\newblock  Phys. Lett. {\bf B 445}  (1998) 178;\relax
H. Baer and J.D. Wells,
\newblock  Phys. Rev. {\bf D 57}  (1998) 4446;\relax
M. Carena, S. Mrenna, C.E.M. Wagner,
\newblock  Phys. Rev. {\bf D 60}  (1999) 075010\relax
\relax
\bibitem{ma30_opal}
OPAL Collaboration, G. Alexander \etal,
\newblock  Z. Phys. {\bf C 73}  (1997) 189\relax
\relax
\end{mcbibliography}

\newpage
\typeout{   }     
\typeout{Using author list for paper 256 -  }
\typeout{$Modified: Jul 15 2001 by smele $}
\typeout{!!!!  This should only be used with document option a4p!!!!}
\typeout{   }
%
%
%
%
%
%

\newcount\tutecount  \tutecount=0
\def\tutenum#1{\global\advance\tutecount by 1 \xdef#1{\the\tutecount}}
\def\tute#1{$^{#1}$}
\tutenum\aachen            
\tutenum\nikhef            
\tutenum\mich              
\tutenum\lapp              
\tutenum\basel             
\tutenum\lsu               
\tutenum\beijing           
\tutenum\berlin            
\tutenum\bologna           
\tutenum\tata              
\tutenum\ne                
\tutenum\bucharest         
\tutenum\budapest          
\tutenum\mit               
\tutenum\panjab            
\tutenum\debrecen          
\tutenum\dublin            
\tutenum\florence          
\tutenum\cern              
\tutenum\wl                
\tutenum\geneva            
\tutenum\hefei             
\tutenum\lausanne          
\tutenum\lyon              
\tutenum\madrid            
\tutenum\florida           
\tutenum\milan             
\tutenum\moscow            
\tutenum\naples            
\tutenum\cyprus            
\tutenum\nymegen           
\tutenum\caltech           
\tutenum\perugia           
\tutenum\peters            
\tutenum\cmu               
\tutenum\potenza           
\tutenum\prince            
\tutenum\riverside         
\tutenum\rome              
\tutenum\salerno           
\tutenum\ucsd              
\tutenum\sofia             
\tutenum\korea             
\tutenum\purdue            
\tutenum\psinst            
\tutenum\zeuthen           
\tutenum\eth               
\tutenum\hamburg           
\tutenum\taiwan            
\tutenum\tsinghua          

{
\parskip=0pt
\noindent
{\bf The L3 Collaboration:}
\ifx\selectfont\undefined
 \baselineskip=10.8pt
 \baselineskip\baselinestretch\baselineskip
 \normalbaselineskip\baselineskip
 \ixpt
\else
 \fontsize{9}{10.8pt}\selectfont
\fi
\medskip
\tolerance=10000
\hbadness=5000
\raggedright
\hsize=162truemm\hoffset=0mm
\def\r{\rlap,}
\noindent

P.Achard\r\tute\geneva\ 
O.Adriani\r\tute{\florence}\ 
M.Aguilar-Benitez\r\tute\madrid\ 
J.Alcaraz\r\tute{\madrid,\cern}\ 
G.Alemanni\r\tute\lausanne\
J.Allaby\r\tute\cern\
A.Aloisio\r\tute\naples\ 
M.G.Alviggi\r\tute\naples\
H.Anderhub\r\tute\eth\ 
V.P.Andreev\r\tute{\lsu,\peters}\
% ...
\newpage
\begin{list}{A}{\itemsep=0pt plus 0pt minus 0pt\parsep=0pt plus 0pt minus 0pt
                \topsep=0pt plus 0pt minus 0pt}
\item[\aachen]
 I. Physikalisches Institut, RWTH, D-52056 Aachen, Germany$^{\S}$\\
 III. Physikalisches Institut, RWTH, D-52056 Aachen, Germany$^{\S}$
\item[\nikhef] National Institute for High Energy Physics, NIKHEF, 
     and University of Amsterdam, NL-1009 DB Amsterdam, The Netherlands
\item[\mich] University of Michigan, Ann Arbor, MI 48109, USA
\item[\lapp] Laboratoire d'Annecy-le-Vieux de Physique des Particules, 
     LAPP,IN2P3-CNRS, BP 110, F-74941 Annecy-le-Vieux CEDEX, France
\item[\basel] Institute of Physics, University of Basel, CH-4056 Basel,
     Switzerland
\item[\lsu] Louisiana State University, Baton Rouge, LA 70803, USA
\item[\beijing] Institute of High Energy Physics, IHEP, 
  100039 Beijing, China$^{\triangle}$ 
\item[\berlin] Humboldt University, D-10099 Berlin, Germany$^{\S}$
\item[\bologna] University of Bologna and INFN-Sezione di Bologna, 
     I-40126 Bologna, Italy
\item[\tata] Tata Institute of Fundamental Research, Mumbai (Bombay) 400 005, India
\item[\ne] Northeastern University, Boston, MA 02115, USA
\item[\bucharest] Institute of Atomic Physics and University of Bucharest,
     R-76900 Bucharest, Romania
\item[\budapest] Central Research Institute for Physics of the 
     Hungarian Academy of Sciences, H-1525 Budapest 114, Hungary$^{\ddag}$
\item[\mit] Massachusetts Institute of Technology, Cambridge, MA 02139, USA
\item[\panjab] Panjab University, Chandigarh 160 014, India.
\item[\debrecen] KLTE-ATOMKI, H-4010 Debrecen, Hungary$^\P$
\item[\dublin] Department of Experimental Physics,
  University College Dublin, Belfield, Dublin 4, Ireland
\item[\florence] INFN Sezione di Firenze and University of Florence, 
     I-50125 Florence, Italy
\item[\cern] European Laboratory for Particle Physics, CERN, 
     CH-1211 Geneva 23, Switzerland
\item[\wl] World Laboratory, FBLJA  Project, CH-1211 Geneva 23, Switzerland
\item[\geneva] University of Geneva, CH-1211 Geneva 4, Switzerland
\item[\hefei] Chinese University of Science and Technology, USTC,
      Hefei, Anhui 230 029, China$^{\triangle}$
\item[\lausanne] University of Lausanne, CH-1015 Lausanne, Switzerland
\item[\lyon] Institut de Physique Nucl\'eaire de Lyon, 
     IN2P3-CNRS,Universit\'e Claude Bernard, 
     F-69622 Villeurbanne, France
\item[\madrid] Centro de Investigaciones Energ{\'e}ticas, 
     Medioambientales y Tecnol\'ogicas, CIEMAT, E-28040 Madrid,
     Spain${\flat}$ 
\item[\florida] Florida Institute of Technology, Melbourne, FL 32901, USA
\item[\milan] INFN-Sezione di Milano, I-20133 Milan, Italy
\item[\moscow] Institute of Theoretical and Experimental Physics, ITEP, 
     Moscow, Russia
\item[\naples] INFN-Sezione di Napoli and University of Naples, 
     I-80125 Naples, Italy
\item[\cyprus] Department of Physics, University of Cyprus,
     Nicosia, Cyprus
\item[\nymegen] University of Nijmegen and NIKHEF, 
     NL-6525 ED Nijmegen, The Netherlands
\item[\caltech] California Institute of Technology, Pasadena, CA 91125, USA
\item[\perugia] INFN-Sezione di Perugia and Universit\`a Degli 
     Studi di Perugia, I-06100 Perugia, Italy   
\item[\peters] Nuclear Physics Institute, St. Petersburg, Russia
\item[\cmu] Carnegie Mellon University, Pittsburgh, PA 15213, USA
\item[\potenza] INFN-Sezione di Napoli and University of Potenza, 
     I-85100 Potenza, Italy
\item[\prince] Princeton University, Princeton, NJ 08544, USA
\item[\riverside] University of Californa, Riverside, CA 92521, USA
\item[\rome] INFN-Sezione di Roma and University of Rome, ``La Sapienza",
     I-00185 Rome, Italy
\item[\salerno] University and INFN, Salerno, I-84100 Salerno, Italy
\item[\ucsd] University of California, San Diego, CA 92093, USA
\item[\sofia] Bulgarian Academy of Sciences, Central Lab.~of 
     Mechatronics and Instrumentation, BU-1113 Sofia, Bulgaria
\item[\korea]  The Center for High Energy Physics, 
     Kyungpook National University, 702-701 Taegu, Republic of Korea
\item[\purdue] Purdue University, West Lafayette, IN 47907, USA
\item[\psinst] Paul Scherrer Institut, PSI, CH-5232 Villigen, Switzerland
\item[\zeuthen] DESY, D-15738 Zeuthen, Germany
\item[\eth] Eidgen\"ossische Technische Hochschule, ETH Z\"urich,
     CH-8093 Z\"urich, Switzerland
\item[\hamburg] University of Hamburg, D-22761 Hamburg, Germany
\item[\taiwan] National Central University, Chung-Li, Taiwan, China
\item[\tsinghua] Department of Physics, National Tsing Hua University,
      Taiwan, China
\item[\S]  Supported by the German Bundesministerium 
        f\"ur Bildung, Wissenschaft, Forschung und Technologie
\item[\ddag] Supported by the Hungarian OTKA fund under contract
numbers T019181, F023259 and T037350.
\item[\P] Also supported by the Hungarian OTKA fund under contract
  number T026178.
\item[$\flat$] Supported also by the Comisi\'on Interministerial de Ciencia y 
        Tecnolog{\'\i}a.
\item[$\sharp$] Also supported by CONICET and Universidad Nacional de La Plata,
        CC 67, 1900 La Plata, Argentina.
\item[$\triangle$] Supported by the National Natural Science
  Foundation of China.
\end{list}
}
\vfill


\newpage

\begin{table}
\begin{center}
\begin{tabular}{|c|ccccc|}
\hline
Scenario & $\mathrm{M_{SUSY}}$ & $\mathrm{M_2}$ & $\mu$ & $\mathrm{X_t}$ & $\mathrm{M_3}$ \\
         & [GeV] & [GeV] & [GeV] & [GeV] & [GeV] \\ 
\hline
``$\mh-$max''    & 1000 & 200     & \phantom{0}$-200$ & 2$\mathrm{M_{SUSY}}$ & 800 \\
``no mixing''    & 1000 & 200     & \phantom{0}$-200$ & \phantom{$-$30}0                    & 800 \\
``large$-\mu$''  & \phantom{0}400 & 400 & \phantom{$-$}1000 & $-$300                 & 200 \\ 
\hline
\end{tabular}
\end{center}
\caption{Values of the MSSM parameters for the three 
scenarios considered in this Letter.
}
\label{tab:mssm_param}
\end{table}

\begin{table}
\begin{center}
\begin{tabular}{|l|cc|c|}
\hline
\multicolumn{4}{|c|}{``$\mh-$max'' scenario} \\
\hline
($\mA$,$\mh$) [GeV] &\multicolumn{2}{|c|}{(90,90)}& (165,110) \\
 \tanb  &\multicolumn{2}{|c|}{25} & 3 \\
\hline
 \multicolumn{4}{|c|}{Preselection}\\
\hline
Data       & \multicolumn{3}{|c|}{2096} \\
Background & \multicolumn{3}{|c|}{2044} \\
\cline{2-4} 
$\mathrm{\epsilon(hA\ra b\bar{b}b\bar{b})}$ &\multicolumn{2}{|c|}{\phantom{.}94\%}& \phantom{.4}$-$ \\
$\mathrm{\epsilon(hZ\ra b\bar{b}q\bar{q})}$ &\multicolumn{2}{|c|}{\phantom{.}89\%}& 92\%\\
hA signal &\multicolumn{2}{|c|}{\phantom{0}7.2}& \phantom{.4}$-$    \\
hZ signal &\multicolumn{2}{|c|}{0.66}& 18.4 \\  
\hline
 \multicolumn{4}{|c|}{Selection}\\
\hline
Analysis branch & \phantom{.}hA & \phantom{.}hZ & \phantom{.}hZ \\
\hline
Data       & \phantom{.9}25   & \phantom{.}275   &  \phantom{.}121   \\
Background & 28.9 & \phantom{.}259 &  \phantom{.}120 \\
$\mathrm{\epsilon(hA\ra b\bar{b}b\bar{b})}$  & 56\% & 19\% & \phantom{.0}$-$ \\
$\mathrm{\epsilon(hZ\ra b\bar{b}q\bar{q})}$  & 25\% & 37\% & 63\% \\
hA signal & \phantom{0}4.3 & \phantom{0}1.5 & \phantom{.0}$-$    \\
hZ signal & 0.18 & 0.28 & 12.6 \\
\hline
 \multicolumn{4}{|c|}{$s/b$ $>$ 0.05} \\
\hline
Data       &\multicolumn{2}{|c|}{\phantom{.0}18}&  \phantom{.0}42   \\
Background &\multicolumn{2}{|c|}{21.3}&  42.1 \\
$\mathrm{\epsilon(hA\ra b\bar{b}b\bar{b})}$ & \multicolumn{2}{|c|}{56\%}& \phantom{.0}$-$\\
$\mathrm{\epsilon(hZ\ra b\bar{b}q\bar{q})}$ & \multicolumn{2}{|c|}{28\%}&55\%\\
hA signal & \multicolumn{2}{|c|}{\phantom{0}4.3}  & \phantom{.0}$-$    \\
hZ signal & \multicolumn{2}{|c|}{0.21} & 11.0 \\  
\hline
\end{tabular} 
\end{center}
\caption{The number of data, expected background, expected signal 
events and signal efficiencies in the four-jet channel after preselection,
after final selection and after applying a cut on the 
final discriminants corresponding to a signal-to-background ratio 
greater than 0.05. This cut is used to calculate the
confidence levels. Numbers are given for two Higgs 
mass hypotheses in the ``$\mh-$max'' scenario.}
\label{tab:bbbb}
\end{table}

\begin{table}
\begin{center}
\begin{tabular}{|l|c|c|}
\hline
\multicolumn{3}{|c|}{``$\mh-$max'' scenario} \\
\hline
($\mA$,$\mh$) [GeV]  & (90,90)  
                     & (165,110) \\
 \tanb               & 25  
                     &  3 \\
\hline
\multicolumn{3}{|c|}{Selection} \\
\hline
Data        & \phantom{.0}28   & \phantom{.00}8    \\
Background  & 27.4 & \phantom{0}6.4 \\
$\mathrm{\epsilon(hA\ra b\bar{b}\tau^+\tau^-)}$ & 41\% & \phantom{.0}$-$ \\ 
$\mathrm{\epsilon(hA\ra \tau^+\tau^- b\bar{b})}$ & 41\% & \phantom{.0}$-$ \\
$\mathrm{\epsilon(hZ\ra b\bar{b}\tau^+\tau^-)}$ & 22\% & 33\% \\
$\mathrm{\epsilon(hZ\ra \tau^+\tau^- q\bar{q})}$ & 21\% & 32\% \\
hA signal & 0.50 & \phantom{.0}$-$    \\
hZ signal & 0.03 & 0.78 \\
\hline
\multicolumn{3}{|c|}{$s/b$ $>$ 0.05} \\
\hline
Data        & \phantom{10.}2    & \phantom{10.}3   \\
Background  & \phantom{0}1.6    & \phantom{0}2.8 \\
$\mathrm{\epsilon(hA\ra b\bar{b}\tau^+\tau^-)}$  & 34\% & \phantom{.0}$-$ \\ 
$\mathrm{\epsilon(hA\ra \tau^+\tau^- b\bar{b})}$ & 30\% & \phantom{.0}$-$ \\
$\mathrm{\epsilon(hZ\ra b\bar{b}\tau^+\tau^-)}$  & 19\% & 30\% \\
$\mathrm{\epsilon(hZ\ra \tau^+\tau^- q\bar{q})}$ & 14\% & 29\% \\
hA signal & 0.39 & \phantom{.0}$-$     \\
hZ signal & 0.02 & 0.70  \\
\hline
\end{tabular}
\end{center}
\caption{
The number of data, expected background, expected signal 
events and signal efficiencies in the channels containing
tau leptons after selection and after applying a cut on the 
final discriminants corresponding to a signal-to-background ratio 
greater than 0.05. This cut is used to calculate the
confidence levels. Numbers are given for two Higgs 
mass hypotheses in the ``$\mh-$max'' scenario.}
\label{tab:bbtt}
\end{table}

\newpage

\end{document}